\documentclass[titlepage,preprint,amsmath,amssymb,aps,superscriptaddress]{revtex4-2}

\usepackage{graphicx}
\usepackage{hyperref}
\usepackage{bm}
\usepackage{dcolumn}
\usepackage{tabstackengine}
\usepackage[dvipsnames]{xcolor}
\usepackage{ulem}
\stackMath{}

\newcommand{\dd}[2]{\frac{\mathrm{d}#1}{\mathrm{d} #2}} %Full first derivative
\newcommand{\pd}[2]{\frac{\partial #1}{\partial #2}} %Partial first derivative
\newcommand{\pdd}[2]{\frac{\partial^2 #1}{\partial #2^2}} %Partial second derivative

\begin{document}

\title{Fluid-fluid phase separation in a soft porous medium}

\author{Oliver W. Paulin}\affiliation{Department of Engineering Science, University of Oxford, Oxford, OX1 3PJ, UK}
\author{Liam C. Morrow}\affiliation{Department of Engineering Science, University of Oxford, Oxford, OX1 3PJ, UK}
\author{Matthew~G.~Hennessy}\affiliation{Department of Engineering Mathematics, University of Bristol, Bristol, BS8 1TW, UK}\affiliation{Mathematical Institute, University of Oxford, Oxford, OX2 6GG, UK}
\author{Christopher W. MacMinn}\email{christopher.macminn@eng.ox.ac.uk}\affiliation{Department of Engineering Science, University of Oxford, Oxford, OX1 3PJ, UK}

\begin{abstract}
Various biological and chemical processes lead to the nucleation and growth of non-wetting fluid bubbles within the pore space of a granular medium, such as the formation of gas bubbles in liquid-saturated lake-bed sediments. In sufficiently soft porous materials, the non-wetting nature of these bubbles can result in the formation of open cavities within the granular solid skeleton. Here, we consider this process through the lens of phase separation, where thermomechanics govern the separation of the non-wetting phase from a fluid--fluid--solid mixture. We construct a phase-field model informed by large-deformation poromechanics, in which two immiscible fluids interact with a poroelastic solid skeleton. Our model captures the competing effects of elasticity and fluid--fluid--solid interactions. We use a phase-field damage model to capture the mechanics of the granular solid. As a model problem, we consider an initial distribution of non-wetting fluid in the pore space that separates into multiple cavities. We use simulations and linear-stability analysis to identify the key parameters that control phase separation, the conditions that favour the formation of cavities, and the characteristic size of the resulting cavities.\end{abstract}

% \keywords{<porous material, elastic material, granular material, phase separation}

\date{\today}

\maketitle

\section{Introduction}

Multiphase flow through rigid porous materials has been studied extensively, but multiphase flow through \textit{deformable} porous materials remains relatively poorly understood \cite{Juanes2020}. These systems are challenging because flows through deformable porous media inherently induce a strong two-way coupling: flow can deform the pore structure, which in turn affects the flow. Recent works have explored this coupling by considering the injection of an immiscible fluid into a saturated, deformable porous medium \cite{Holtzman2012,Meng2020,Carrillo2021a,Carillo2021b}, highlighting how capillary forces can deform the structure of the pore space. Similar phenomena have also attracted attention from the perspective of pattern formation in multiphase frictional flows \cite{Sandnes2011,Dumazer2016a}. 

\begin{figure*}
    \centering
    \includegraphics{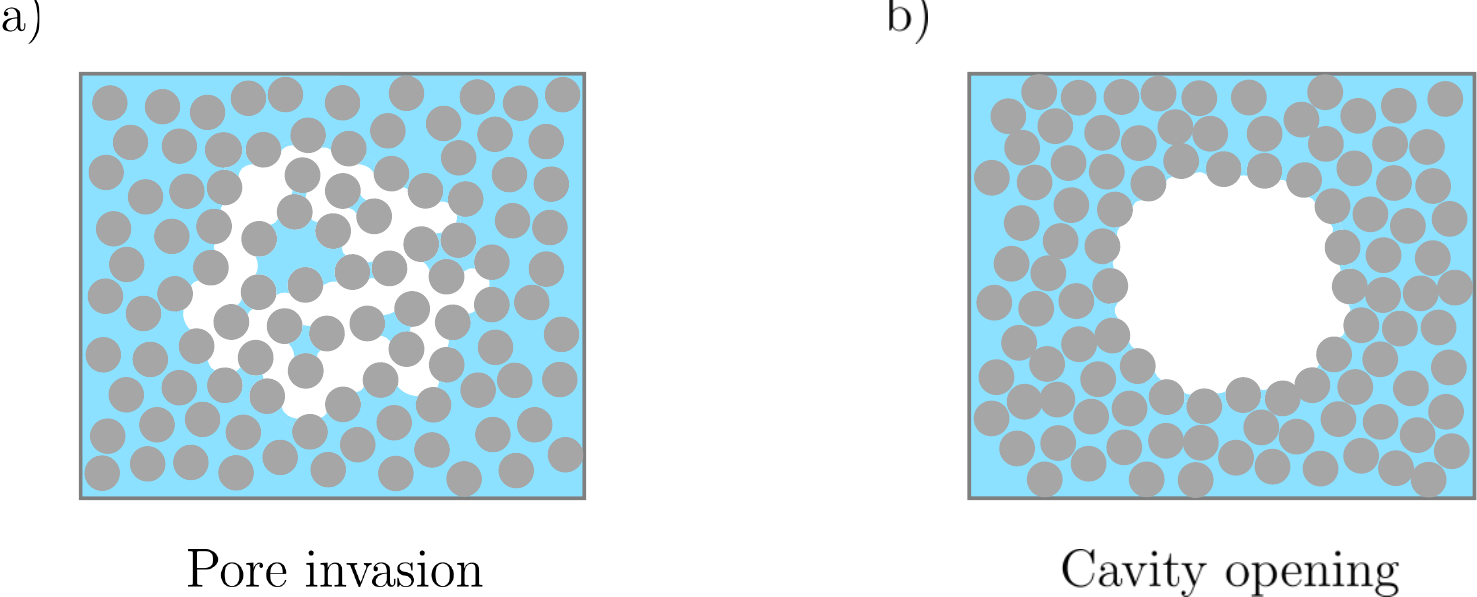}
    \caption{A schematic showing the two different states of non-wetting-phase occupancy within a soft porous medium with a granular microstructure: (a) the non-wetting phase (white) and the wetting phase (blue) can coexist within the pore space, between the solid grains (grey) or (b) the non-wetting phase can displace the solid grains to form an open cavity. Cavity opening occurs when the solid skeleton is soft enough for capillary forces to overcome the resistance to deformation (\textit{e.g.}, elasticity).}
    \label{fig:Schematic}
\end{figure*}

Deformation enables the emergence of completely new flow phenomena that do not have a rigid analogue, the most striking of which is the ability of the non-wetting phase to form open (solid-free) pathways and cavities \cite{Wheeler1988,Boudreau2005}. Non-wetting cavities form due to capillary forces, which make it energetically costly for the non-wetting phase to invade narrow pore throats. If the solid skeleton is sufficiently soft, it may be energetically favourable for the non-wetting phase to instead displace solid and form macroscopic cavities within the porous skeleton \cite{Wheeler1988,Boudreau2005}. The non-wetting phase can thus occupy one of two distinct states within a soft porous medium; either (1) invading the pore space by displacing the wetting phase or (2) forming open cavities by additionally displacing the solid (Figure~\ref{fig:Schematic}). 
In a soft porous medium, the latter is resisted by elasticity and the competition between capillarity and elasticity determines which of these states is energetically favourable. The pore-scale physics of cavity formation are relatively well understood \cite{Jain2009,Sills1991}, but a continuum-scale model for this phenomenon is currently lacking. A quantitative understanding of the continuum-scale physics of cavity formation would have application to a variety of natural and industrial systems, including, for example, predicting the rate of gas venting from lake beds, sea beds and waste ponds \cite{Skarke2014b,Kam2001,Lee2020}. 

The spontaneous formation of non-wetting cavities within a soft porous medium can be conceptualised as a process of thermodynamic phase separation. Recently, the problem of liquid-liquid phase separation within an elastic network has attracted substantial attention from the perspective of droplet formation within living cells \cite{Hyman2014,Style2018,Rosowski2020,FernandezRico2022,Wei2020,Kothari2020}. Elastically modulated phase separation has also been studied within the context of hydrogels \cite{Onuki1999,Zhou2010,Hennessy2020,Celora2022}. It is now well understood that the presence of an elastic network fundamentally alters the phase-separation process by introducing an energetic cost to the rearrangements that are needed for phase separation to occur. This energetic cost limits the final size of phase-separated domains \cite{Wei2020}, decreases the rate of post-separation coarsening \cite{Onuki1999}, and in some cases can suppress phase separation entirely \cite{Wei2020,Kothari2020}. Fluid-fluid phase separation in a poroelastic medium has many similarities to these systems, as demonstrated and discussed below.

Phase-field models provide a powerful framework for modelling phase-separation processes. These models were first developed within the materials-science community to describe spinodal decomposition during the solidification of alloys \cite{CahnHilliard1958,Cahn1961,Provatas2010}. The resulting Cahn-Hilliard equation has since been generalised for use in a range of problems in fluid dynamics \citep[\textit{e.g.,}][]{Anderson1998b,Cueto-Felgueroso2012c,Cueto-Felgueroso2014b,Fu2016a}, including multiphase flow in \textit{rigid} porous media \cite{Cueto-Felgueroso2008,Schreyer2021}. Many Cahn--Hilliard-style models focus on two phases, either one fluid and one solid \citep[\textit{e.g.,}][]{Hennessy2020}, two fluids \citep[\textit{e.g.,}][]{Fu2017a}, or two solids \citep[\textit{e.g.,}][]{Larche1992}, but these models have also been generalised to an arbitrary number of phases \cite{Kim2005b}. Although most Cahn--Hilliard-style models are derived via an ad-hoc variational approach, Gurtin \cite{Gurtin1989b,Gurtin1995,Gurtin1996a} later developed a thermodynamically consistent framework for these models by introducing the concept of subscale `microforces' associated with changes in local composition. Gurtin's framework also allows for the natural inclusion of a deformable solid phase via conservative (elastic) and/or dissipative (plastic) contributions to the free energy \cite{Gurtin1996a}, which are otherwise challenging to capture consistently in a traditional Cahn-Hilliard model.

Here, we develop a continuum model that captures the fundamental mechanics of the formation of non-wetting cavities within a soft porous medium. By conceptualising the transition between pore invasion and cavity formation as a phase-separation process, our phase-field model captures the competing physical processes in a thermodynamically consistent manner. In \S \ref{Model section}, we derive our model by extending the work of Hennessy \textit{et al.} \cite{Hennessy2020} to include an additional fluid phase and to allow for elastic degradation of the solid skeleton via a damage model \cite{Miehe2010b,Tang2019}. Our derivation follows the approach of Gurtin \cite{Gurtin1996a}, using balance laws to set out an energy inequality that imposes certain restrictions on the allowable constitutive behaviours. In \S \ref{2 phase section}, we simplify our model to one spatial dimension and benchmark it against known solutions for two different two-phase limits. In \S \ref{3 phase section}, we consider a nearly uniform initial distribution of gas within the pore space of a slightly compressed solid skeleton, which may or may not lead to the spontaneous formation of multiple gas cavities. We then study the onset of phase separation through numerical simulations (\S \ref{Numerical Simulations}) and  linear stability analysis (\S \ref{LSA section}). In \S \ref{Cavity Size}, we investigate the characteristic size of the resulting cavities. Finally, in \S \ref{Conclusions section}, we summarise our findings and highlight areas for future work.

\section{Model Development} \label{Model section}

We begin by considering a mixture of three phases, two fluids and a solid. At the pore scale, these phases exist within separate domains separated by sharp interfaces. At the continuum (Darcy) scale, however, the phases are mixed and the proportion of each at a particular point in space and time is measured by its volume fraction. The volume fractions thus act as phase-field order parameters that can be used to distinguish the physical characteristics of different regions, and to smoothly interpolate between them. Relative motion of the phases is governed by a thermodynamic framework that drives the system toward minima of its total free energy, as described in detail below.

We take the solid phase to be a porous skeleton comprising non-cohesive, incompressible solid `grains'. The pore space between the grains is occupied by two immiscible, incompressible fluid phases. For clarity and without loss of generality, we refer to these fluids as `liquid' and `gas', where the gas is the non-wetting phase. The key feature of our model is that, under certain conditions, it is energetically favourable for the gas to separate from the other two phases by forming open cavities. 

We next formulate our model. We begin with the kinematics of the mixture (\S \ref{Kinematics}). We then formulate balance laws for mass and momentum (\S \ref{balance laws}) and consider the thermodynamics of the mixture (\S \ref{Thermodynamics}). We formulate a free energy that captures the energetic characteristics of our system (elasticity, damage, and gas-liquid-solid interactions; \S \ref{Free Energy}). We close the model by using thermodynamic arguments to determine constitutive relationships that link the various quantities in our balance laws (\S \ref{Constitutive Behaviour}). Finally, we simplify our model to the case of a one dimensional (1D) system (\S \ref{Uniaxial}).

\subsection{Kinematics} \label{Kinematics}

Fluid-flow problems are typically considered in an Eulerian reference frame (fixed in space), whereas solid-mechanics problems are typically considered in a Lagrangian reference frame (fixed to the solid material). For a poromechanical system, however, the fluids and the solid co-exist and must be considered in the same reference frame. To do so, it is convenient to begin by defining both frames and the relationship between them.

The deformation of the solid skeleton is defined by comparing its current (deformed) state to an undeformed (relaxed) reference state. The Lagrangian coordinate~$\mathbf{X}$ refers to the reference state and the Eulerian coordinate~$\mathbf{x}$ refers to the current state. We denote spatial gradients with respect to the Lagrangian and Eulerian coordinates by $\nabla_0$ and $\nabla$, respectively. The Lagrangian displacement of the solid is then $\mathbf{U}_s\left(\mathbf{X},t\right)=\mathbf{x}\left(\mathbf{X},t\right)-\mathbf{X}$ and the Eulerian displacement is $\mathbf{u}_s \left(\mathbf{x},t\right)= \mathbf{x} - \mathbf{X}\left(\mathbf{x},t\right)$. The deformation gradient tensor, $\mathbb{F}$, describes the deformation of the solid skeleton relative to its reference state, \begin{equation}\label{FDefinition}
    \mathbb{F} = \nabla_0\mathbf{x}.
\end{equation}
Note that we use the convention $\left(\nabla\mathbf{A}\right)_{ij}=\pd{A_i}{x_j}$ for a general vector field $\mathbf{A}=A_i\mathbf{\hat{e}}_i$ and $\left(\nabla\cdot\mathbb{B}\right)_i=\pd{B_{ij}}{x_j}$ for a general tensor field  $\mathbb{B}=B_{ij}\mathbf{\hat{e}}_i\mathbf{\hat{e}}_j$, where $\mathbf{\hat{e}}_i$ and $\mathbf{\hat{e}}_j$ are Cartesian unit vectors and where we have adopted the Einstein summation convention. Using the definitions above, $\mathbb{F}$ is related to $\mathbf{U}_s$ and $\mathbf{u}_s$ via
\begin{equation}\label{F in u}
    \mathbb{F} = \mathbb{I}+\nabla_0\mathbf{U}_s = \left[\mathbb{I}-\nabla\mathbf{u}_s\right]^{-1},
\end{equation}
where $\mathbb{I}$ is the identity tensor.

The volume fraction of each phase can be measured relative to either the relaxed or deformed configurations. The nominal volume fractions, $\varphi_{\alpha}$, measure the volume of phase $\alpha$ per unit reference bulk volume, whereas the true volume fractions, $\phi_{\alpha}$, measure the volume of phase $\alpha$ per unit current bulk volume, where $\alpha = s,~l,~g$ indicates the solid, liquid, and gas phases, respectively. These volume fractions are related by $\varphi_{\alpha}=\phi_{\alpha}J$, where the Jacobian determinant, $J = \textrm{det}(\mathbb{F})$, measures the ratio of current (deformed) bulk volume to reference (relaxed) bulk volume and is thus equal to one in the reference state. The no-void condition can thus be expressed in two ways,
\begin{equation}\label{No void condition}
    \sum_{\alpha} \varphi_{\alpha}=J \quad \textrm{and} \quad \sum_{\alpha} \phi_{\alpha}=1.
\end{equation}
A result of the no-void condition is that any two of the three volume fractions fully determine the local composition. 

The incompressibility of the solid grains requires that the nominal volume fraction of solid must remain unchanged during deformation, such that $\varphi_s\equiv\varphi_s^0\equiv\phi_s^0$, where $\varphi_s^0\equiv\phi_s^0$ is the solid fraction in the reference state, which we take to be uniform for simplicity. As a result,
\begin{equation}\label{Incompressibility}
    J=\frac{\phi_s^0}{\phi_s}.
\end{equation}
It is also convenient to define the true porosity, $\phi = \phi_g + \phi_l = 1 - \phi_s$, and the relaxed porosity, $\phi_0=1-\phi_s^0$. Below, we develop evolution equations for $\phi_g$ and $\phi$ and then evaluate $\phi_l=\phi-\phi_g$ and $\phi_s=1-\phi$ as needed. 

\subsection{Balance Laws}\label{balance laws}

We formulate evolution equations for $\phi_g$ and $\phi$ by first considering conservation of mass for each phase. In the Eulerian frame, conservation of mass can be written
\begin{equation}
\label{conservation of mass}
\pd{}{t}\left(\rho_{\alpha}\phi_{\alpha}\right) + \nabla\cdot \left(\rho_{\alpha}\phi_{\alpha}\mathbf{v}_{\alpha}\right) = 0,
\end{equation}
where $\rho_{\alpha}$ and $\mathbf{v}_{\alpha}$ are the density and local velocity of phase $\alpha$, respectively. We assume for simplicity that all three phases are incompressible, meaning that $\rho_\alpha$ is a constant. The Eulerian volume flux, $\mathbf{w}_\alpha$, of phase $\alpha$ relative to the solid skeleton is
\begin{equation}\label{fluid fluxes}
    \mathbf{w}_{\alpha} = \phi_{\alpha}\left( \mathbf{v}_{\alpha}-\mathbf{v}_s\right).
\end{equation}
We reduce Equations~(\ref{conservation of mass}) to two evolution equations, one each for $\phi_g$ and $\phi$, and an elliptic equation for the total mixture flux, $\mathbf{q} = \sum\limits_\alpha \phi_{\alpha}\mathbf{v}_{\alpha} = \mathbf{v}_s + \mathbf{w}_g + \mathbf{w}_l$, by eliminating the phase velocities in favour of the relative fluxes. The result of these manipulations is
\begin{subequations} \label{Collected evolution}
\begin{align}
    \label{Gas evolution}
    \frac{\partial\phi_g}{\partial t} + \nabla\cdot\left(\phi_g\mathbf{q}\right) + \nabla\cdot\left[ \left(1 - \phi_g\right) \mathbf{w}_g - \phi_g\mathbf{w}_l \right]&=0,\\
    \label{Porosity evolution}
   \frac{\partial\phi}{\partial t} + \nabla\cdot\left(\phi\mathbf{q}\right) + \nabla\cdot\left[ \left( 1 -\phi\right) \left(\mathbf{w}_g + \mathbf{w}_l\right)\right]&=0,\\
   \label{Q equation}
    \nabla\cdot\mathbf{q} &= 0.
\end{align}
\end{subequations}
We derive thermodynamically consistent constitutive expressions for the fluxes $\mathbf{w}_{\alpha}$ in \S \ref{Constitutive Behaviour}. We express mass conservation in Lagrangian form by using the Reynolds transport theorem to rewrite Equation (\ref{conservation of mass}) as 
\begin{equation}\label{Lagrangian Mass Consevration}
\pd{}{t}\left(\rho_{\alpha}\varphi_{\alpha}\right) + \nabla_0\cdot \left(\rho_{\alpha}\mathbf{W}_{\alpha}\right) = 0,
\end{equation}
where the nominal volume flux, $\mathbf{W}_{\alpha}$, is related to the true flux by $\mathbf{W}_{\alpha}=J\mathbb{F}^{-1}\mathbf{w}_{\alpha}$.

Further restrictions can be imposed on our system by considering conservation of momentum. The true (Cauchy) total stress, $\mathbb{T}$, measures the total internal force per unit current area within the mixture. In the absence of body forces and neglecting inertia, mechanical equilibrium requires that $\mathbb{T}$ must be divergence free,
\begin{equation}\label{momentumconservation}
    \nabla\cdot\mathbb{T} = 0.
\end{equation}
In the Lagrangian frame, momentum balance is most naturally expressed in terms of the first Piola-Kirchhoff stress, $\mathbb{S}$,
\begin{equation}\label{momentumconservationLagrangian}
\nabla_0\cdot\mathbb{S} = 0,
\end{equation}
where $\mathbb{S}=J\mathbb{T}\mathbb{F}^{-\intercal}$ measures the total internal force per unit reference area within the mixture. 

\subsection{Thermodynamics} \label{Thermodynamics}

The first law of thermodynamics requires that the rate of change of free energy in a particular control volume, via mass flux into the volume or working due to internal compositional changes, cannot exceed the rate of work done by external forces. Following Gurtin \cite{Gurtin1996a}, we formulate this restriction by considering the change in Helmholtz free energy, $\psi$, of a bulk material element in the Lagrangian frame. The rate of change of energy per unit mass of phase $\alpha$ due to a net flux of phase $\alpha$ is measured by the chemical potential, $\mu_{\alpha}$. The key ingredient in Gurtin's approach is that changes in energy resulting from changes in composition, as measured by changes in $\varphi_\alpha$, can be represented by the action of a nominal vector `microstress', $\boldsymbol{\xi}_{\alpha}$, acting on the boundary of the material element. This microstress is intended to capture the net work done by subscale physical mechanisms such as capillary forces that would otherwise not be resolved at the continuum scale. Mechanical damage to the solid skeleton can be captured in the same way by introducing a damage parameter, $d$, and an associated microstress, $\boldsymbol{\xi}_d$ (see \S \ref{Free Energy} below). The working of external forces is represented by the action of the total stress, $\mathbb{S}$, on the boundary of the material element. 
Summing these different contributions leads to a dissipation inequality,
\begin{multline}\label{Gurtin integral}
    \frac{\mathrm{d}}{\mathrm{d} t}\int_{\mathcal{V}_0} \psi \mathrm{d}V \leq 
    -\int_{\partial \mathcal{V}_0} \rho_g\mu_g \mathbf{W}_g \cdot \hat{\mathbf{n}} \mathrm{d}A
    -\int_{\partial \mathcal{V}_0} \rho_l\mu_l \mathbf{W}_l \cdot \hat{\mathbf{n}} \mathrm{d}A 
    +\int_{\partial \mathcal{V}_0} \left(\boldsymbol{\xi}_g \cdot \hat{\mathbf{n}}\right)\dot{\varphi_g} \mathrm{d}A \\
    +\int_{\partial \mathcal{V}_0} \left(\boldsymbol{\xi}_l \cdot \hat{\mathbf{n}}\right)\dot{\varphi_l} \mathrm{d}A 
    +\int_{\partial \mathcal{V}_0} \left(\boldsymbol{\xi}_s \cdot \hat{\mathbf{n}}\right) \dot{J} \mathrm{d}A
    +\int_{\partial \mathcal{V}_0} \left(\boldsymbol{\xi}_d \cdot \hat{\mathbf{n}}\right)\dot{d} \mathrm{d}A
    +\int_{\partial \mathcal{V}_0} \left(\mathbb{S} \cdot \hat{\mathbf{n}}\right) \cdot \dot{\mathbf{u}}_s \mathrm{d}A,
\end{multline}
where $\mathcal{V}_0$ and $\partial\mathcal{V}_0$ represent the material element and its boundary, respectively, and $\dot{\square}~\equiv~\frac{\partial}{\partial t}\square$. Grouping the surface integrals in Equation (\ref{Gurtin integral}) and then using the divergence theorem leads to a local form of this inequality,
\begin{equation}\label{Gurtin local}
\dot{\psi} + \nabla_0\cdot\left(\rho_g\mu_g \mathbf{W}_g + \rho_l\mu_l \mathbf{W}_l
-\dot{\varphi_g} \boldsymbol{\xi}_g 
-\dot{\varphi_l} \boldsymbol{\xi}_l 
-\dot{J} \boldsymbol{\xi}_s 
-\dot{d} \boldsymbol{\xi}_d
-\mathbb{S} \cdot \dot{\mathbf{u}}_s\right) \leq 0.
\end{equation}
We must also ensure that the no-void condition (Equation \ref{No void condition}) is satisfied. This constraint can be incorporated into the above thermodynamic restriction through the use of a Lagrange multiplier, $p$, which is the thermodynamic mixture pressure. To make this constraint compatible with the dimensions of the dissipation inequality, we differentiate Equation (\ref{No void condition}) with respect to time and use Jacobi's formula to evaluate the derivative of $J$, arriving at
\begin{equation}\label{Incompressibility differentiated}
    \dot{J} = J \mathbb{F}^{-\intercal}\colon\dot{\mathbb{F}} = \dot{\varphi_g} + \dot{\varphi_l}.
\end{equation}
We combine the dissipation inequality (Equation \ref{Gurtin local}) with the incompressibility constraint (Equation \ref{Incompressibility differentiated}) by multiplying the latter by $p$ and summing the two. 

Finally, we elucidate the consequences of this inequality by asserting that $\psi$ can, in general, be a function of composition via $\varphi_g$, $\varphi_l$, $d$, and their gradients, and of deformation via $\mathbb{F}$ and $\nabla_0 J$, so that $\psi~=~\psi\left(\varphi_g, \varphi_l, d, \mathbb{F}, \nabla_0\varphi_g, \nabla_0\varphi_l, \nabla_0 d, \nabla_0 J \right)$. Note that we could equivalently write $\psi$ as a function of true volume fractions and their Eulerian gradients, but it is convenient to use nominal quantities as the independent variables here. We then use the chain rule as well as conservation of mass (Equation \ref{Lagrangian Mass Consevration}) and conservation of momentum (Equation \ref{momentumconservationLagrangian}) to arrive at
\begin{equation}\label{Gurtin collected}
\begin{split}
  &\left(\frac{\partial\psi}{\partial\varphi_g} - \rho_g\mu_g +p -\nabla_0\cdot\boldsymbol{\xi}_g\right)\dot{\varphi_g}+
  \left(\frac{\partial\psi}{\partial\varphi_l} - \rho_l\mu_l +p -\nabla_0\cdot\boldsymbol{\xi}_l\right)\dot{\varphi_l}\\&+
  \left(\frac{\partial\psi}{\partial\nabla_0\varphi_g} -\boldsymbol{\xi}_g\right) \nabla_0\dot{\varphi_g}+
  \left(\frac{\partial\psi}{\partial\nabla_0\varphi_l} -\boldsymbol{\xi}_l\right) \nabla_0\dot{\varphi_l}+
  \left(\frac{\partial\psi}{\partial d} - \nabla_0\cdot\boldsymbol{\xi}_d\right)\dot{d}\\&+
  \left(\frac{\partial\psi}{\partial\nabla_0 d} -\boldsymbol{\xi}_d\right) \nabla_0\dot{d}+
  \left(\frac{\partial\psi}{\partial\mathbb{F}} - \mathbb{S} - pJ\mathbb{F}^{-\intercal} -J\left(\nabla_0\cdot\boldsymbol{\xi}_s\right)\mathbb{F}^{-\intercal} \right)\colon\dot{\mathbb{F}}\\&+
  \left(\frac{\partial\psi}{\partial\nabla_0 J} - \boldsymbol{\xi}_s \right) \nabla_0\dot{J}+
  \mathbf{W}_g\cdot\nabla_0\left(\rho_g\mu_g\right) 
  + \mathbf{W}_l\cdot\nabla_0\left(\rho_l\mu_l\right) \leq 0.
\end{split}
\end{equation}
Due to the mutual independence of the arguments of $\psi$, this inequality is only satisfied for all possible configurations if each of the bracketed terms vanishes independently. This requirement then leads to a set of thermodynamic constraints,
\begin{subequations}\label{constraints}
\begin{equation}\label{phig constraint}
   \rho_g\mu_g= p+ \pd{\psi}{\varphi_g}-\nabla_0\cdot\pd{\psi}{\nabla_0\varphi_g},
\end{equation}
\begin{equation}\label{phil constraint}
    \rho_l\mu_l= p+\pd{\psi}{\varphi_l}-\nabla_0\cdot\pd{\psi}{\nabla_0\varphi_l},
\end{equation}
\begin{equation}\label{d constraint}
     \pd{\psi}{d}-\nabla_0\cdot\pd{\psi}{\nabla_0 d}=0,
\end{equation}
\begin{equation}\label{S constraint}
    \mathbb{S}=-pJ\mathbb{F}^{-\intercal}+\pd{\psi}{\mathbb{F}}-J\left(\nabla_0\cdot\pd{\psi}{\nabla_0 J}\right)\mathbb{F}^{-\intercal},
\end{equation}
\end{subequations}
and the additional requirement that
\begin{equation}
    \label{Flux Condition}
    \mathbf{W}_g\cdot\nabla_0\left(\rho_g\mu_g\right) + \mathbf{W}_l\cdot\nabla_0\left(\rho_l\mu_l\right) \leq 0.
\end{equation}
The latter expression can be satisfied by choosing an appropriate form for the fluxes $\mathbf{W}_g$ and $\mathbf{W}_l$, as discussed in \S \ref{Constitutive Behaviour} below.

\subsection{Free Energy} \label{Free Energy}

The evolution equations derived in \S \ref{balance laws} drive the system toward an equilibrium state defined by the minima of $\psi$. We must next construct a function $\psi$ that captures the different physical processes at play in our system. We assume that $\psi$ is, in general, an additive function of each physical contribution,
\begin{equation}\label{Psi All}
    \psi= \psi_{\mathrm{bulk}} +\psi_{\mathrm{mix}} + \psi_{\mathrm{elastic}} + {\psi_{\mathrm{damage}}} + \psi_{\mathrm{interface}}, 
\end{equation}
where $\psi_{\mathrm{bulk}}$, $\psi_{\mathrm{mix}}$, $\psi_{\mathrm{elastic}}$, ${{\psi_{\mathrm{damage}}}}$ and $\psi_{\mathrm{interface}}$ are the energetic contributions from changing the mass of gas or liquid in the material element, gas-liquid-solid interactions within the material element, elasticity of the solid skeleton, mechanical damage to the solid skeleton, and interfacial effects, respectively. We write these contributions in terms of the true volume fractions and their Eulerian gradients in order to provide greater physical insight; conversion to nominal quantities is straightforward.

The bulk free energy, $\psi_{\mathrm{bulk}}$, is the free energy associated with the amount of gas and liquid in the material element and is given by
\begin{equation}\label{Psi1}
    \psi_{\mathrm{bulk}}\left(\phi_\alpha,J\right) = \left(\mu_l^0\rho_l\phi_l + \mu_g^0\rho_g\phi_g \right)J,
\end{equation}
where the reference chemical potential $\mu^0_\alpha$ is the energy associated with adding one unit mass of fluid $\alpha$ to the material element, neglecting interactions between phases. For immiscible materials, $\mu^0_\alpha$ is a constant. 

{The mixing (interaction) energy, $\psi_{\mathrm{mix}}$, depends on the pore-scale arrangement of the phases and on the wetting characteristics of the fluid-fluid-solid system. To construct a free energy that gives the desired phenomenological behaviour for our system, we draw inspiration from Flory-Huggins theory used in polymer physics \cite{Flory1943}. In a polymer solution, the Flory-Huggins energy, $\psi_{\mathrm{FH}}$, for a multiphase mixture is
\begin{equation}\label{FHEnergy}
    \psi_{\mathrm{FH}}=k_B TJ\left[\sum_{\alpha\neq s} \Omega_\alpha\phi_\alpha\log\phi_\alpha+\sum_{\alpha\neq\beta}\hat{\chi}_{\alpha\beta}\phi_\alpha\phi_\beta\right],
\end{equation}
where $k_B$ is the Boltzmann constant, $T$ is temperature, $\Omega_\alpha$ is inversely proportional to the characteristic size of particles of phase $\alpha$, and $\hat{\chi}_{\alpha\beta}$ characterises the strength of unfavourable interactions between phases $\alpha$ and $\beta$. The first sum in equation (\ref{FHEnergy}) represents the entropy of mixing of the different phases. Due to the large size of solid grains compared to fluid molecules, it is assumed that $\Omega_s\ll\Omega_{f}$, where the subscript $s$ indicates a general solid phase and $f$ a general fluid phase, and so the contribution of solid to the entropy term is neglected. The entropy term promotes the mixing of different phases, and also constrains the relevant volume fractions to remain between zero and one. The second sum in equation (\ref{FHEnergy}) represents the enthalpy of interactions between different phases, and results in the formation of double-well potentials which promote phase separation. The Flory-Huggins free energy thus captures the key features we expect from gas-liquid-solid interactions in a granular system, namely a double-well structure in which volume fractions are constrained to remain between zero and one. As such, we use $\psi_{\mathrm{FH}}$ as motivation for constructing an appropriate form for $\psi_{\mathrm{mix}}$.}

{In our system, the most important interaction is that the gas is non-wetting to the solid relative to the liquid. We enforce this requirement by assuming that gas-liquid and gas-solid interactions are much more energetically expensive than liquid-solid interactions, meaning that $\hat{\chi}_{ls}\ll\hat{\chi}_{gs,gl}$. We therefore neglect the liquid-solid interaction. In general, the remaining coefficients, $\hat{\chi}_{gs}$ and $\hat{\chi}_{gl}$, depend on physical quantities such as surface energies and pore structure; we take them to be constant material properties here. For simplicity, we further assume that $\Omega_g=\Omega_l=\Omega$. The mixing energy is then given by
\begin{equation} \label{Psi3}
    \psi_{\mathrm{mix}}\left(\phi_\alpha,J\right) = \mathcal{E}_{\mathrm{mix}}\left(\phi_g\log{\phi_g} + \phi_l\log{\phi_l} + \chi_{gs}\phi_g\phi_s + \chi_{gl}\phi_g\phi_l\right)J,
\end{equation}
where $\chi_{\alpha\beta}=\frac{\hat{\chi}_{\alpha\beta}}{\Omega}$ and $\mathcal{E}_{\mathrm{mix}}$ is a characteristic energy density associated with mixing, which we treat as a material property, and into which we have absorbed $\Omega$. Figure \ref{fig:TernaryPlot} shows a ternary plot of this interaction energy. The key feature of the energetic landscape is that there are local minima near $\phi_g=1$ and along $\phi_g=0$, corresponding to open gas cavities and to a liquid-saturated porous matrix, respectively. 
\begin{figure}
    \centering
    \includegraphics{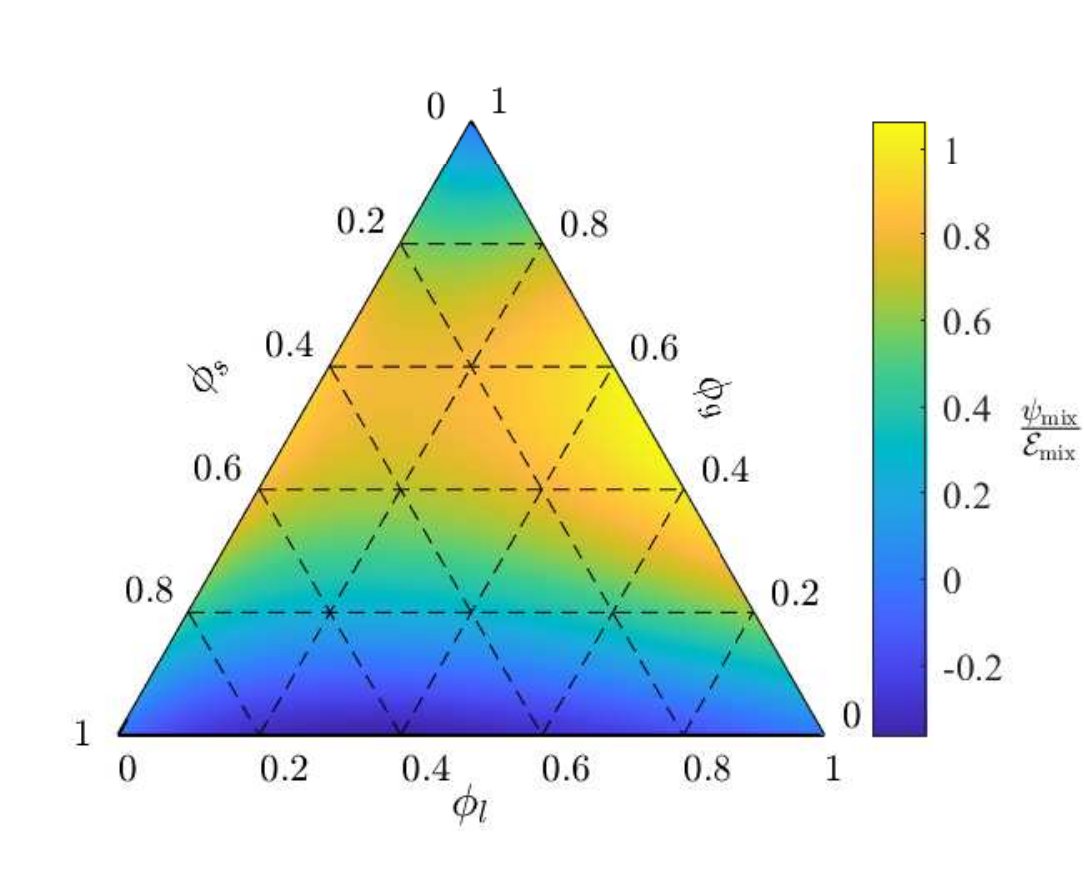}
    \caption{An example ternary plot of $\frac{\psi_{\mathrm{mix}}}{\mathcal{E}_{\mathrm{mix}}}$ (Equation \ref{Psi3}) with local minima at $\phi_g=1$ (gas cavities) and along $\phi_g=0$ (liquid-saturated porous medium). For this figure, we set $\chi_{gs}=5$ and $\chi_{gl}=7$.}
    \label{fig:TernaryPlot}
\end{figure}
The specific values of $\chi_{gs}$ and $\chi_{gl}$ control the relative depth of the two minima, but not the overall structure of the energy landscape.}

The elastic energy of the solid matrix due to deformation is given by $\psi_{\mathrm{elastic}}$, and is usually expressed in terms of a strain-energy density function, $\mathcal{W}_{\mathrm{el}}\left(\mathbb{F}\right)$. For simplicity, we use a standard neo-Hookean strain-energy density,
\begin{equation}\label{NeoHookean 3D}
    \mathcal{W}_{\mathrm{el}} = \frac{1}{2}\mathcal{G}\left[\textrm{tr}\left(\mathbb{F}^\intercal\mathbb{F}\right)-3-2\log{J}\right]+\frac{1}{2}\mathcal{K}\left(\log{J}\right)^2,
\end{equation}
where $\mathcal{G}$ and $\mathcal{K}$ are the `drained' shear and bulk moduli of the solid skeleton, respectively. Note that the derivation that follows is valid for any choice of $\mathcal{W}_{\mathrm{el}}$. 

A notable characteristic of a solid skeleton with a granular microstructure is that tension does not result in the storage of elastic energy, since grains only interact when in contact. In order to specify a different elastic response between open cavities and a continuous solid packing, we define an additional order parameter, $d\left(\mathbf{x},t\right)$, which we identify as a Bourdin-style damage parameter \cite{Bourdin2008}. The damage parameter represents a phase-field approximation of the fractures to the solid skeleton that are caused by the opening of gas cavities. Locally, a value of $d=0$ corresponds to an undamaged solid skeleton, in which the solid grains are in contact and hence provide a standard elastic response to compression. In contrast, a value of $d=1$ corresponds to a fully damaged solid skeleton that provides no elastic resistance to further compression or tension. The use of a phase-field damage parameter ensures a smooth transition between open cavities and the rest of the porous medium 

To implement damage, we assume that the strain-energy density can be decomposed into distinct tensile and compressive parts, denoted $\mathcal{W}^+$ and $\mathcal{W}^-$, respectively \cite{Miehe2010b,Tang2019}. We demonstrate a suitable decomposition of $\mathcal{W}_{\mathrm{el}}$ into tensile and compressive parts for a non-linear material in Appendix \ref{Detailed Mechanics}, following the approach of Tang \textit{et al.} \cite{Tang2019}. We can then assume that the tensile part of the elastic energy is degraded by $d$ \cite{Miehe2010b}, such that,
\begin{subequations}\label{Psi2}
\begin{equation}
    \psi_{\mathrm{elastic}}\left(d,\mathbb{F},J\right) = g\left(d\right) \mathcal{W}^++\mathcal{W}^-,
\end{equation}
\begin{equation}\label{Degradation}
    g\left(d\right)=\left(1-\Theta\right)\left(1-d\right)^2+\Theta,
\end{equation}
\end{subequations}
where $\Theta\ll 1$ is a numerical smoothing parameter. In the undamaged limit ($d=0$), this formulation reduces to $\psi_{\mathrm{elastic}}=\mathcal{W}^++\mathcal{W}^-=\mathcal{W}_{\mathrm{el}}$, as expected.

{According to Griffith's theory of fracture, the energy required to create a fracture in a material is given by the critical fracture energy, $\mathcal{G}_c$, integrated over the area of the fracture \cite{Griffiths1921}. In a granular medium, the `fracture' energy is associated with the energy required to cause the decohesion of neighbouring solid grains. In our phase-field approach, the energy required to form such fractures is approximated by a bulk energy, $\psi_{\mathrm{damage}}$, \cite{Miehe2010b,Bourdin2008}, of the form,
\begin{equation}\label{Psifrac}
    \psi_{\mathrm{damage}}\left(d,\nabla_0d\right)=\mathcal{G}_c\left(\frac{d^2}{4l_d}+l_d\vert\nabla_0d\vert^2\right),
\end{equation}
where $l_d$ controls the width of the transition between damaged and undamaged material. For a non-cohesive solid skeleton, $\mathcal{G}_c$ is very small, with the only resistance to decohesion coming from wetting liquid bridges between solid grains. Note also that damage of a non-cohesive solid skeleton is reversible: if a cavity closes, $d$ will return to zero and the skeleton will return to standard elastic behaviour.}

Finally, $\psi_{\mathrm{interface}}$ measures the energy of diffuse interfaces between different phases. In a phase-field formulation, interfaces are represented implicitly as regions with large gradients in composition. The interfacial energy thus depends on gradients in volume fraction \cite{CahnHilliard1958,Hong2013} and the total interfacial energy is the sum of the energy of each of the possible types of interface \cite{Kim2005b}. The simplest such dependence can in general be written as
\begin{equation}\label{psi int general}
    \Upsilon_\textrm{int}=\sum_\alpha\frac{\gamma_\alpha}{2} |\nabla \phi_\alpha|^2 J,
\end{equation}
where $\gamma_\alpha$ are the associated interfacial coefficients. For a two-phase system (say, liquid and gas), the no-void condition allows elimination of one of the two volume fractions (say, $\phi_g$) such that $\Upsilon_{\textrm{int}}=\frac{\gamma}{2}|\nabla\phi_l|^2J$, where $\gamma=\gamma_l+\gamma_g$, and $\gamma$ can be directly related to the surface energy of the sharp interface formed between these two phases. For our three-phase system, we use the no-void condition to eliminate $\phi_s$ and write the total interfacial energy solely in terms of the gas and liquid fractions,
\begin{equation} \label{Psi4}
    \psi_{\mathrm{interface}}\left(J,\nabla\phi_\alpha\right) = \frac{\gamma_1}{2} |\nabla \phi_g|^2J + \frac{\gamma_2}{2} |\nabla \phi_l|^2J + \frac{\gamma_3}{2} \left(\nabla \phi_g\cdot\nabla \phi_l\right) J,
\end{equation}
where $\gamma_1=\gamma_g+\gamma_s$, $\gamma_2=\gamma_l+\gamma_s$ and $\gamma_3=2\gamma_s$. Unlike for a two-phase system, the relationship between these quantities and the associated surface energies is not clear.

\subsection{Constitutive Behaviour} \label{Constitutive Behaviour}

The free energy constructed in \S \ref{Free Energy} can now be combined with the thermodynamic constraints resulting from the energy inequality (Equations \ref{constraints}) to find expressions for the constitutive behaviour of the system in terms of the different driving mechanisms. We use Equations (\ref{phig constraint}) and (\ref{phil constraint}) to decompose the chemical potential for the gas and liquid phases as the sum of the pressure and a capillary potential, $\Psi_{\alpha}$,
\begin{subequations}\label{chem pot all}
\begin{align}
    \label{Chemical potential}
      \rho_{\alpha}\mu_\alpha &= p + \Psi_{\alpha},\\
      \label{Capillary definition}
    \Psi_{\alpha} &=  \frac{\partial\psi}{\partial\varphi_\alpha} - \nabla_0\cdot\left(\frac{\partial\psi}{\partial\nabla_0\varphi_\alpha}\right) \equiv \frac{\delta \psi}{\delta \varphi_{\alpha}}.
\end{align}
\end{subequations}
The capillary potentials represent the thermodynamic forcing on the gas and liquid phases due to the interactions between them. Note that $\Psi_\alpha$ can be viewed as the variational derivative of $\psi$ with respect to $\varphi_\alpha$.

Equation (\ref{d constraint}) leads to an elliptic equation for $d$ as a function of the tensile elastic energy. For our specified free energy, the result is
\begin{equation}\label{d elliptic}
    4l_d\left(1-\Theta\right)\left(1-d\right)\mathcal{W}^++4\mathcal{G}_cl_d^2\nabla_0^2d=\mathcal{G}_c d.
\end{equation}
{In the limit $\mathcal{G}_c\to0$ (no energetic resistance to decohesion), Equation (\ref{d elliptic}) reduces to $\mathcal{W}^+\left(1-d\right)=0$, such that any non-zero tension results in immediate fracture of the solid skeleton. This case essentially reproduces the sharp-interface limit of open cavities within a poroelastic continuum.}

Equation (\ref{S constraint}) allows the total stress to be decomposed into three distinct parts, converting from the first Piola-Kirchhoff stress to the Cauchy stress to arrive at
\begin{equation}\label{Stress decomposition}
    \mathbb{T} = -p\mathbb{I} + \boldsymbol{\sigma}' + \mathbb{K}.
\end{equation}
The effective (elastic) stress, $\boldsymbol{\sigma}'$, is
\begin{equation}\label{Stress definition}
    \boldsymbol{\sigma}'=\frac{1}{J}\frac{\partial\psi_{\mathrm{elastic}}}{\partial\mathbb{F}}\mathbb{F}^{\intercal}=g\left(d\right)\frac{1}{J}\frac{\partial\mathcal{W}^+}{\partial\mathbb{F}}\mathbb{F}^{\intercal}+\frac{1}{J}\frac{\partial\mathcal{W}^-}{\partial\mathbb{F}}\mathbb{F}^{\intercal},
\end{equation}
where we have used $\psi_\mathrm{elastic}$ from \S \ref{Free Energy} and $g\left(d\right)$ is the degradation function defined in Equation (\ref{Degradation}). The Korteweg stress is
\begin{equation}\label{Korteweg Definition}
     \mathbb{K}=\frac{1}{J}\frac{\partial\psi_{\mathrm{interface}}}{\partial\mathbb{F}}\mathbb{F}^{\intercal}- \left(\nabla_0\cdot\frac{\partial\psi_{\mathrm{interface}}}{\partial\nabla_0 J}\right)\mathbb{I}.
\end{equation}
The Korteweg stress is the bulk representation of interfacial tension at diffuse, internal interfaces and thus resists gradients in composition. The stress balance derived in Equation (\ref{momentumconservation}) allows us to link the effective and Korteweg stresses with gradients in the chemical potential via the pressure field.

Recall that Equation (\ref{Flux Condition}) constrains the relationship between fluid chemical potentials and fluxes. The simplest way to satisfy this constraint is for the fluxes to be proportional to gradients in chemical potential, such that,
\begin{equation}
    \mathbf{W}_{\alpha}=-\mathbb{M}_\alpha\nabla_0\left(\rho_{\alpha}\mu_{\alpha}\right),
\end{equation}
where $\mathbb{M}_\alpha\left(\varphi_{\alpha},\mathbb{F}\right)$ is a positive definite mobility tensor. The associated Eulerian volume fluxes are given by
\begin{equation}\label{Darcy-like}
      \mathbf{w}_{\alpha}=-\mathcal{M}_\alpha\nabla\left(\rho_{\alpha}\mu_\alpha\right),
\end{equation}
where $\mathcal{M}_\alpha\left(\phi_{\alpha}\right)$ is the Eulerian mobility of fluid $\alpha$. In the Eulerian frame, the pore space of granular packings is typically fairly isotropic, so we assume a scalar mobility for simplicity. At the least, $\mathcal{M}_\alpha$ must be a linear function of the fluid volume fraction, so that if there is no fluid present, then there will be no associated flux. In the interest of developing a simple model that captures the leading-order behaviour of the system, we choose $\mathcal{M}_\alpha=\phi_\alpha \mathcal{M}_\alpha^0$, where $\mathcal{M}_\alpha^0$ is a constant for a fluid $\alpha$ within a particular porous medium; however, more complicated mobility laws could also be used. For example, $\mathcal{M}_\alpha$ could be decomposed into a product of viscosity, absolute permeability, and relative permeability. These relations between the fluid fluxes and their respective chemical potentials complete our model. 

Explicit expressions for the effective stress, the capillary potential and the Korteweg stress for the particular form of the free energy specified in \S \ref{Free Energy} are given in Appendices \ref{Detailed Mechanics}, \ref{Capillary Potential} and \ref{Korteweg Stress}, respectively. A full set of governing equations is presented in Appendix \ref{Full Model Appendix}.

\subsection{Reduction to 1D}\label{Uniaxial}

\begin{figure}
    \centering
    \includegraphics[width=8.6cm]{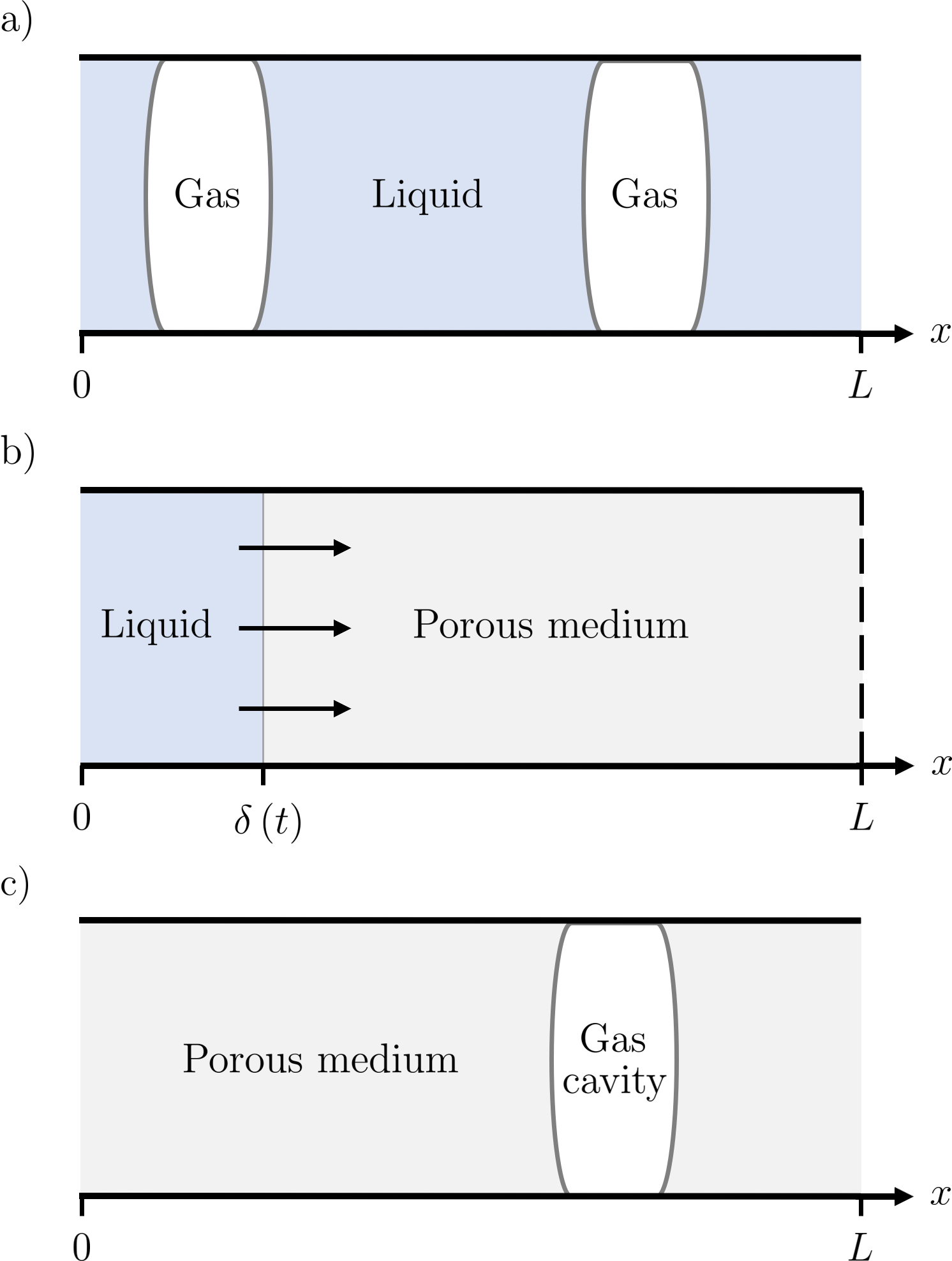}
    \caption{A schematic showing the uniaxial (1D) set up for each of the three model problems considered below: (a) the separation of gas and liquid in the absence of a solid skeleton in a periodic domain, (b) the deformation of a saturated porous medium via an applied liquid flow, in the absence of gas, and (c) the formation of gas cavities in a periodic porous domain. All flow and deformation take place in the $x$ direction.}
    \label{fig:Uniaxial}
\end{figure}

For uniaxial flow and deformation, the model derived above can be reduced to one dimension. Although uniaxial behaviour is a strong constraint, studying such a system is significantly easier, both conceptually and computationally, and allows us to gain substantial insight into what is ultimately a complex model. Uniaxial behaviour implies that
\begin{subequations}
\begin{equation}
    \mathbf{v}_{\alpha} = v_{\alpha}\left(x,t\right) \mathbf{\hat{e}}_x,
\end{equation}
\begin{equation}
    \mathbf{u}_s = u_s\left(x,t\right) \mathbf{\hat{e}}_x,
\end{equation}
\begin{equation}
    \phi_{\alpha} = \phi_{\alpha}\left(x,t\right),
\end{equation}
\begin{equation}
    d = d\left(x,t\right),
\end{equation}
\end{subequations}
as illustrated in Figure \ref{fig:Uniaxial}.

At this point, it is convenient to non-dimensionalise our model. We do so using the following scalings:
$x=L\tilde{x}$, $q=q_0\tilde{q}$, and $t=\tau\tilde{t}$, where $L$ is a characteristic length scale, $q_0$ is the mixture flux at the boundary, and $\tau=\frac{L^2}{\mathcal{M}^0_l\mathcal{E}_\textrm{mix}}$ is the time scale associated with the transport of liquid across a distance $L$ by capillary effects. This scaling motivates us to introduce the following dimensionless groups, which characterise the relative strengths of the different physical processes at play:
\begin{align*}
    Q_0=\frac{q_0\tau}{L},~~
    \mathcal{M}_r=\frac{\mathcal{M}^0_g}{\mathcal{M}^0_l},\quad
    \mathcal{D}=\frac{\mathcal{E}_{\mathrm{mix}}}{\mathcal{G}},\quad
    \Gamma_i=\frac{\gamma_i}{\mathcal{E}_{\mathrm{mix}}L^2},\quad
    \kappa_c=\frac{\mathcal{G}_c}{\mathcal{E}_\mathrm{mix}l_d},\quad
    \Lambda_d=\frac{l_d^2}{L^2}.
\end{align*}
The strength of background fluid flow relative to capillary forces is measured by $Q_0$, while $\mathcal{M}_r$ is the ratio of gas mobility to liquid mobility, $\mathcal{D}$ is the characteristic `deformability' of the solid, $\Gamma_i$ ($i=1,2,3$) are the rescaled interfacial coefficients, $\kappa_c$ is the strength of cohesion between solid grains, and $\Lambda_d$ characterises the sharpness of the transition between damaged and undamaged material. We work exclusively with dimensionless quantities going forward, dropping the tildes for convenience.

In 1D, Equation (\ref{Q equation}), which enforces the incompressibility of the mixture, simplifies to
\begin{equation}\label{QND}
\frac{\partial q}{\partial x} = 0.
\end{equation}
For the scalings defined above, the non-dimensional boundary condition for the flux is $q\left(0,t\right)=Q_0$. The solution to Equation (\ref{QND}) is therefore given by $q=Q_0$. {For uniaxial flow and deformation, the deformation gradient tensor, $\mathbb{F}$, is diagonal, which greatly simplifies the stresses: only the $xx$-components of $\boldsymbol{\sigma}'$ and $\mathbb{K}$, denoted ${\sigma}'_{xx}$ and $K_{xx}$, respectively, appear explicitly in the 1D model.} We use momentum conservation (Equation \ref{momentumconservation}) and the stress decomposition (Equation \ref{Stress decomposition}) to link $p$ with $\sigma'_{xx}$ and $K_{xx}$:
\begin{equation}
    \pd{p}{x}=\pd{}{x}\left({\sigma}'_{xx}+K_{xx}\right).
\end{equation}
We now replace the fluid fluxes in the 1D evolution equation for $\phi_g$, given by Equation (\ref{Gas evolution}), with the constitutive behaviour outlined above in Equations (\ref{chem pot all}) and (\ref{Darcy-like}). The gas fraction then evolves according to
\begin{equation}\label{PhiGND}
\begin{split}
    \frac{\partial\phi_g}{\partial t} = 
    -Q_0\pd{\phi_g}{x}
    +\mathcal{M}_r&\pd{}{x}\left[
    \left(1-\phi_g\right) \phi_g \frac{\partial}{\partial x} \left( 
    {\Psi}_g
    + {\sigma}'_{xx}
    + {K}_{xx}
   \right)\right]\\
   -& \pd{}{x}\left[ \phi_g\left(\phi-\phi_g\right) \frac{\partial}{\partial x}\left(
    {\Psi}_l
    + {\sigma}'_{xx}
    + {K}_{xx}
       \right)
\right],
\end{split}
\end{equation}
where $\Psi_\alpha$ is now the dimensionless capillary potential, which is specified below. Similarly, we combine the 1D evolution equation for $\phi$ (Equation \ref{Porosity evolution}) with the relevant constitutive behaviour to give
\begin{equation}
    \begin{split}
        \label{PhiND}
    \frac{\partial\phi}{\partial t} = 
    -Q_0\pd{\phi}{x}
    +\mathcal{M}_r&\pd{}{x}\left[
    \left(1-\phi\right) \phi_g \frac{\partial}{\partial x} \left( 
    {\Psi}_g
    + {\sigma}'_{xx}
    + {K}_{xx}
   \right)\right]\\
+& \pd{}{x}\left[ \left(1-\phi\right)\left(\phi-\phi_g\right) \frac{\partial}{\partial x}\left(
    {\Psi}_l
    + {\sigma}'_{xx}
    + {K}_{xx}
       \right)
\right].
    \end{split}
\end{equation}
We close our model with the 1D version of Equation (\ref{d elliptic}), which describes the distribution of damage,
\begin{equation}\label{d ND}
    4\left(1-\Theta\right)\left(1-d\right){\mathcal{W}}^++4\kappa_c\Lambda_d\pdd{d}{X}=\kappa_c d,
\end{equation}
where ${\mathcal{W}}^+$ is the dimensionless tensile energy, which in 1D is solely a function of $\phi$. The model is thus reduced to two non-linear evolution equations (for $\phi_g$ and for $\phi$) and one elliptic equation (for $d$).

By non-dimensionalising the capillary potentials (Equation \ref{Capillary definition}) and undertaking the required differentiation (Appendix \ref{Capillary Potential}), we find that
\begin{subequations}
\begin{equation}
    {\Psi}_{g} = {\Pi}_g -\Gamma_1\frac{\partial^2 \phi_g}{\partial x^2}
    -\frac{\Gamma_3}{2}\frac{\partial^2 \phi_l}{\partial x^2},
\end{equation}
\begin{equation}
    {\Psi}_{l} = {\Pi}_l -\Gamma_2\frac{\partial^2 \phi_l}{\partial x^2}
    -\frac{\Gamma_3}{2}\frac{\partial^2 \phi_g}{\partial x^2},
\end{equation}
\end{subequations}
where the interaction potentials, ${\Pi}_g$ and ${\Pi}_l$, are
\begin{subequations}
\begin{equation}
{\Pi}_g = \log{\phi_g} + \phi_s + \left(1-\phi_g\right)\left(\chi_{gl}\phi_l + \chi_{gs}\phi_s \right)
\end{equation}
\textrm{and}
\begin{equation}
{\Pi}_l = \log{\phi_l} + \phi_s + \phi_g\left(\chi_{gl}\phi_g - \chi_{gs}\phi_s\right).
\end{equation}
\end{subequations}
The interaction potentials result from the thermodynamic interactions between the phases, as described by $\psi_\textrm{mix}$, and are the drivers for phase separation.
 
For uniaxial deformation, tension can be defined solely volumetrically: the material is in tension if $J>1$ and in compression if $J\leq1$. The non-dimensional effective stress (Equation \ref{Stress definition}) thus reduces to
\begin{equation}\label{stressND}
    {\sigma}'_{xx}=\begin{cases}
    {\sigma}'_0\left(\phi\right) &J\leq1,\\
    {\sigma}'_0\left(\phi\right)  g\left(d\right) &J>1,
    \end{cases}
\end{equation}
where ${\sigma}'_0$ is the undamaged neo-Hookean stress given by
\begin{equation}\label{1D elastic law}
    {\sigma}'_0=\frac{1}{\mathcal{D}}\left[J-\frac{1}{J}+\varsigma\frac{\log{J}}{J}\right],
\end{equation}
and $\varsigma=\frac{\mathcal{K}}{\mathcal{G}}$. The derivation of this effective stress can be found in Appendix \ref{Detailed Mechanics}. Similarly, the tensile energy used in Equation (\ref{d ND}) is given by,
\begin{equation}
    {\mathcal{W}}^+= \begin{cases}
    0 &\textrm{for}~J\leq 1,\\
    {\mathcal{W}}_{el}^0 & \textrm{for}~J>1,
    \end{cases}
\end{equation}
where ${\mathcal{W}}_{el}^0=\frac{1}{2\mathcal{D}}\left[J^2-1-2\log{J}+\varsigma\left(\log{J}\right)^2\right]$. The incompressibility condition (Equation~\ref{Incompressibility}) links $J$ to the porosity field:
\begin{equation}
    J=\frac{1-\phi_0}{1-\phi}.
\end{equation}
Finally, $K_{xx}$ is derived in Appendix \ref{Korteweg Stress} and is given by
\begin{equation}
    \begin{split}
        \label{KortewegND}
{K}_{xx} = 
\Gamma_1\left[\phi_g\frac{\partial^2\phi_g}{\partial x^2}-\frac{1}{2}\left(\frac{\partial\phi_g}{\partial x}\right)^2\right]
&+\Gamma_2\left[\phi_l\frac{\partial^2\phi_l}{\partial x^2}-\frac{1}{2}\left(\frac{\partial\phi_l}{\partial x}\right)^2\right]\\
&+\frac{1}{2}\Gamma_3\left[\phi_g\frac{\partial^2\phi_l}{\partial x^2}+\phi_l\frac{\partial^2\phi_g}{\partial x^2}-\frac{\partial\phi_l}{\partial x}\frac{\partial\phi_g}{\partial x}\right].
    \end{split}
\end{equation}
Equations (\ref{PhiGND})-(\ref{KortewegND}) comprise a complete model for our system in 1D, and can be solved numerically given appropriate initial and boundary conditions. We do so below by discretising in space using finite differences on a staggered grid and integrating in time using MATLAB's built-in solver for stiff differential equations, \texttt{ode15s}. For the remainder of this paper, we limit ourselves to the 1D case. We do not anticipate fundamentally different physical behaviour in 2D or 3D. 

\section{Two-Phase Problems} \label{2 phase section}

Our full three-phase model captures several different physical effects, and ultimately consists of modelling fluid-fluid phase separation within the framework of large-deformation poroelasticity. To gain insight into the behaviour of our model, we begin by considering two limiting cases involving two-phase mixtures: a gas-liquid system (no solid) and a liquid-solid system (no gas). These two limiting cases are comparatively well understood, allowing us to benchmark the predictions of our model against previous results.

\subsection{Gas-Liquid System (no solid)}

In the no-solid limit, $\phi\equiv\phi_0\equiv 1$, our model reduces to a gas-liquid system in which the two phases separate in an unconstrained domain (no porous medium), as illustrated in Figure \ref{fig:Uniaxial}a. This scenario is typically described by the Cahn-Hilliard equation for binary mixtures \cite{CahnHilliard1958}. In this limit, the evolution equation for the porosity (Equation \ref{PhiND}) and the elliptic equation for the damage (Equation \ref{d ND})  degenerate. The evolution equation for the gas fraction (Equation \ref{PhiGND}) simplifies greatly, most notably due to disappearance of the elastic terms, becoming
\begin{equation}\label{CH full}
    \frac{\partial\phi_g}{\partial t}  
    -\mathcal{M}_r\pd{}{x}
    \left[
   \phi_g \frac{\partial \mu_g}{\partial x}     \right]=0.
\end{equation}
The gas chemical potential is now given by
\begin{equation}\label{CH mug}
    \mu_g=\log{\phi_g}+\chi_{gl}\left(1-\phi_g\right)^2-\Gamma\left[\frac{1}{2}\left(\pd{\phi_g}{x}\right)^2+\left(1-\phi_g\right)\pdd{\phi_g}{x}\right],
\end{equation}
where we have defined $\Gamma=\Gamma_1+\Gamma_2-\Gamma_3=\Gamma_g+\Gamma_l$ as the characteristic interfacial coefficient between the fluid phases. Equations (\ref{CH full}) and (\ref{CH mug}) resemble the standard Cahn-Hilliard equation for fluid-fluid phase separation \citep[\textit{e.g.,}][]{Fu2016a} but, in our case, also account for the mechanical stress generated across diffuse interfaces.

\begin{figure}
    \centering
    \includegraphics{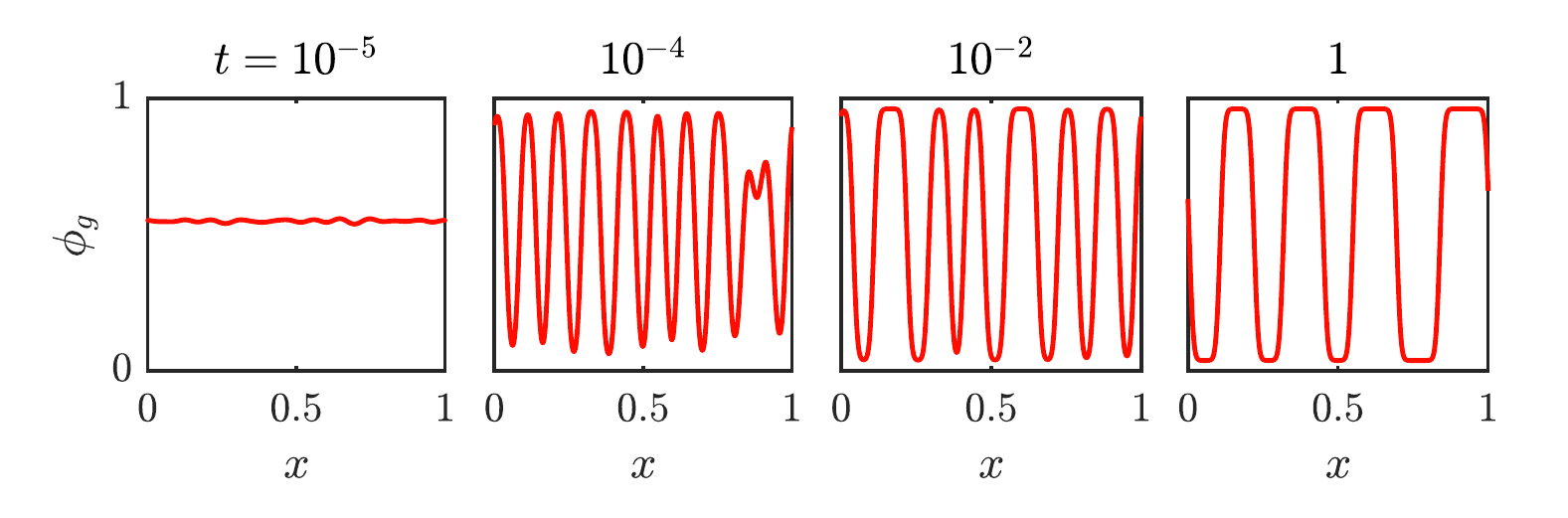}
    \caption{Gas-liquid phase separation via the numerical solution of Equations (\ref{CH full}) and (\ref{CH mug}) subject to periodic boundary conditions. An initially nearly homogeneous mixture undergoes the classical spinodal decomposition and subsequent coarsening characteristic of Cahn--Hilliard-type problems. Parameter values of $\mathcal{M}_r=50$, $\chi_{gl}=3.5$, and $\Gamma=0.0002$ were used, along with an initial gas fraction of $0.55$ with a random perturbation of size $0.01$. The simulations were stopped before reaching the final equilibrium due to the slow rate of coarsening in 1D.}
    \label{fig:CH}
\end{figure}

We now consider a test  problem with periodic boundary conditions. For the initial condition, we use a homogeneous gas fraction superimposed with small, random perturbations. The subsequent evolution is standard fluid-fluid phase separation, also known as spinodal decomposition, in which the mixture separates spontaneously into distinct gas-rich and liquid-rich domains separated by diffuse interfaces (Figure \ref{fig:CH}).
Over time, these domains coarsen as smaller gas `bubbles' merge with larger gas `bubbles'. Coarsening is driven by the evolution of the system toward minimum global energy (the smaller the number of bubbles, the smaller the total interfacial energy). In 1D, coarsening is very slow \cite{Sun2000}; as such, we stop our simulations before reaching the eventual equilibrium of a single bubble.

\subsection{Liquid-Solid System (no gas)}\label{LiquidSolidSection}

In the no-gas limit, our model reduces to a diffuse-interface formulation of the well-known theory of single-phase large-deformation poroelasticity \cite{Biot1972,Coussy2004}. In this case, there will be no thermodynamically driven phase separation because there is minimal energetic cost to liquid remaining within the pore space of the solid skeleton. However, flow can still drive phase separation, as demonstrated below. Setting $\phi_g\equiv0$, Equation (\ref{PhiGND}) for the evolution of the gas fraction degenerates and Equation (\ref{PhiND}) for the evolution of the porosity simplifies to 
\begin{equation}\label{LSCase}
\frac{\partial\phi}{\partial t} +Q_0\frac{\partial \phi}{\partial x}
-\pd{}{x}\left[
    \left(1-\phi\right)\mathcal{M}\left(\phi\right)\frac{\partial}{\partial x}\left(
{\sigma}'_{xx}+{\Psi}
      \right)
    \right]=0,
\end{equation}
with
\begin{equation}
    {\Psi} = 
    \log{\phi} +1-\phi-
\Gamma_2\left[\left(1-\phi\right)\frac{\partial^2\phi}{\partial x^2}+\frac{1}{2}\left(\frac{\partial\phi}{\partial x}\right)^2\right],
\end{equation}
and a dimensionless mobility given by $\mathcal{M}\left(\phi\right)=\phi$. This equation is reminiscent of standard sharp-interface poroelasticity \citep[\textit{e.g.,}][]{MacMinn2016}, but with the addition of a potential, ${\Psi}$, that allows the formation of internal diffuse interfaces, such as those resulting from the fluid-driven opening of solid-free cavities within the porous skeleton. The effective stress here also depends on $d$ via Equation (\ref{d ND}).

For a quantitative comparison of our model with sharp-interface poroelasticity, we consider the uniaxial deformation of a packing of soft grains due to net fluid flow from left to right (Figure \ref{fig:Uniaxial}b). We impose a constant liquid influx at the left boundary of non-dimensional flow rate $Q_0$. We take the right boundary to be permeable, allowing free outflow of the liquid but not of the solid. The result is that the packing will be compressed in the direction of the flow, with the left edge acting as an internal free boundary that moves downstream away from its initial position. Traditionally, this problem would be approached by solving the equations of large-deformation poroelasticity within the packing while explicitly tracking the position of the left edge of the packing as a moving boundary, assuming that the effective stress vanishes there. In our model, the left edge of the packing is an internal diffuse interface between a solid-free cavity ($\phi\sim1$) and a porous domain ($0<\phi<1$).

For this problem, the sharp-interface formulation can be solved analytically at steady state, thus allowing for direct comparison with our phase-field approach. Following MacMinn \textit{et al.} \cite{MacMinn2016}, we define the left edge of the packing to have a position $\delta\left(t\right)$, with a fluid-filled domain for $x<\delta$. According to the theory of large-deformation poroelasticity, the porosity in the domain $x~\in~[\delta\left(t\right),1]$ then evolves according to
\begin{equation}\label{SI poroelasticity}
    \frac{\partial\phi}{\partial t} +Q_0\frac{\partial \phi}{\partial x}
-\pd{}{x}\left[
    \left(1-\phi\right)\mathcal{M}\left(\phi\right)
\pd{{\sigma}'_{0}}{x}
      \right]=0.
\end{equation}
We anticipate that our phase-field model would reduce to this formulation in the sharp-interface limit. Conservation of mass requires that the influx of liquid at the left boundary, $x=\delta$, must match the outflux at the right boundary, $x=1$. We use this fact, along with the requirement that the solid velocity vanishes at $x=1$, to enforce that the relative liquid flux at the right boundary is equal to the total mixture flux, $w_l=-\mathcal{M}\pd{{\sigma}'_{0}}{x}=Q_0$. We also assume that the solid packing is relaxed at its left edge, so that the effective stress vanishes at $x=\delta$. The boundary conditions for Equation (\ref{SI poroelasticity}) are then given by
\begin{equation}\label{SI boundary conditions}
    {\sigma}'_{0}\left(x=\delta\right)=0,
\qquad
    \phi\pd{{\sigma}'_{0}}{x}\Bigr|_{x=1}=-Q_0.
\end{equation}
As discussed above, the steady-state porosity is piecewise:
\begin{equation}\label{piecewise}
    \phi\left(x\right)=\begin{cases}
    1 & x < \delta,\\
    f\left(x\right) & \delta \leq x \leq 1.
        \end{cases}
\end{equation}
Solving Equation (\ref{SI poroelasticity}) subject to the boundary conditions above (Equations \ref{SI boundary conditions}) gives $f(x)$. An analytical solution for $f(x)$ is derived in Appendix \ref{Analytic} and plotted in Figure \ref{fig:FDD}.
\begin{figure}
    \centering
    \includegraphics{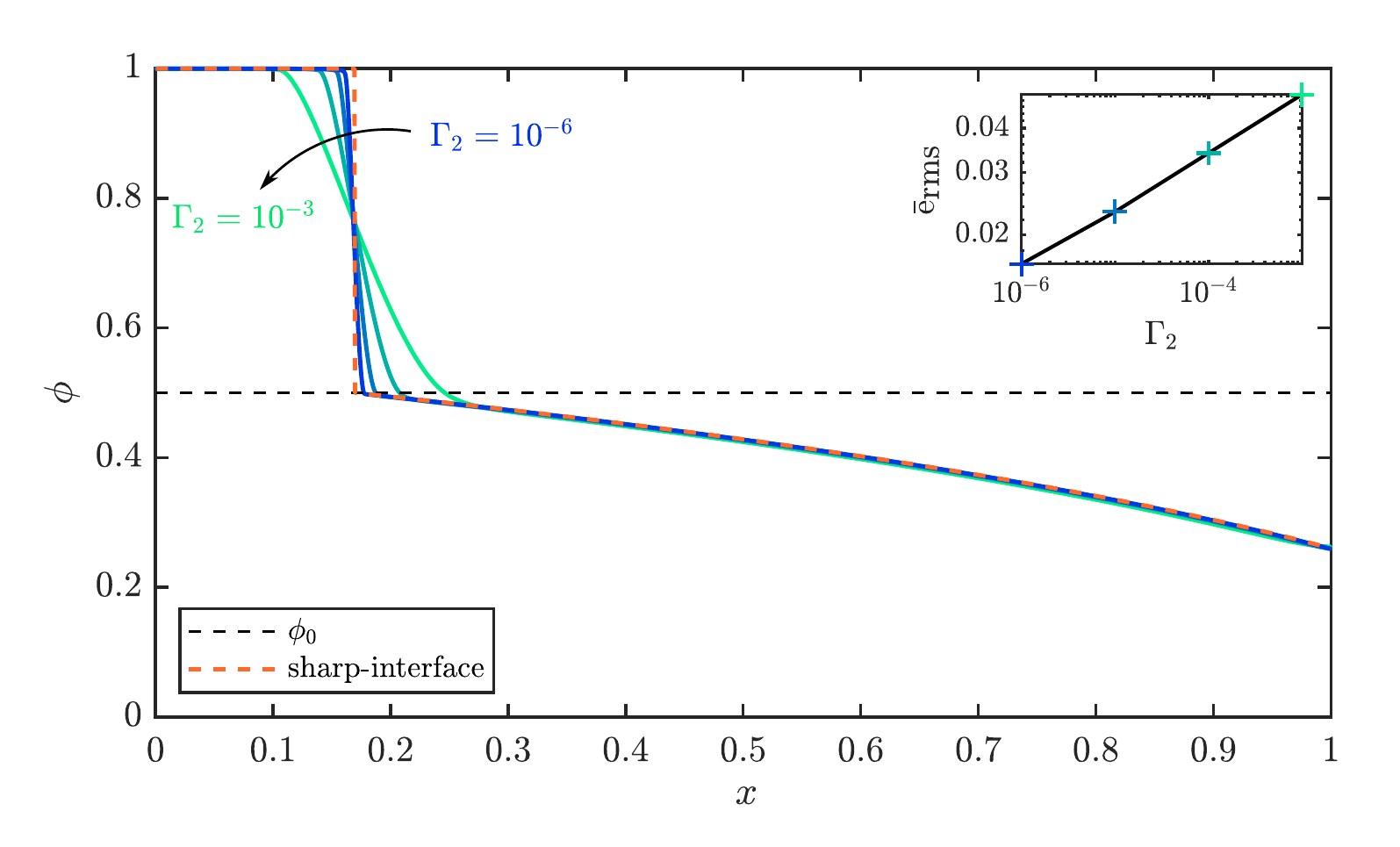}
    \caption{Fluid-driven deformation of a poroelastic medium at steady state from the sharp-interface model (analytical solution, Equation \ref{piecewise}) and from our phase-field formulation (numerical solutions to Equation \ref{LSCase}) for $\Gamma_2=10^{-6}$, $10^{-5}$, $10^{-4}$ and $10^{-3}$ (dark to light colours). The phase-field results approach the sharp-interface result as $\Gamma_2$ decreases. Here, we set $\phi_0=0.5$, $Q_0=0.2$, $\mathcal{D}=2$, $\varsigma=0.1$, $\kappa_c=10^{-8}$ and $\Lambda_d=0.1\Gamma_2$. The inset shows the variation of root mean square (RMS) difference between the sharp-interface and phase-field solutions, $\bar{\textrm{e}}_\textrm{rms}$, with $\Gamma_2$.}
    \label{fig:FDD}
\end{figure}

To compare our phase-field model to the sharp-interface result, we solve Equation (\ref{d ND}) and Equation (\ref{LSCase}) numerically until reaching steady state. With regard to boundary conditions for these equations, we assume that the solid velocity vanishes at each end of the domain, so that the relative liquid flux, $w_l$, is equal to $Q_0$ at each end. To enforce the fact that the solution should not depend on anything exterior to the domain, we also impose that the gradients of porosity and damage vanish at both ends. The boundary conditions for Equation (\ref{LSCase}) are thus
\begin{equation}
    w_l\left(x=0,1\right)=Q_0,
\quad
    \pd{\phi}{x}\Bigr|_{x=0,1}=0,
\quad
    \pd{d}{x}\Bigr|_{x=0,1}=0.
\end{equation}
With these boundary conditions, we solve Equations (\ref{d ND}) and (\ref{LSCase}) numerically with initial condition $\phi\left(x,0\right)=\phi_0$. 

Figure \ref{fig:FDD} compares the steady state of the phase-field simulations for various values of $\Gamma_2$ (solid lines), with the analytical solution to the sharp-interface model given in Appendix~\ref{Analytic} (dashed line). In both cases, the solution comprises a solid-free region at the left and a compressed porous packing at the right. Within the latter, the porosity is non-linear with position. Away from the interface, the two solutions are in excellent qualitative and quantitative agreement. Near the interface, the discontinuity in the sharp-interface solution is approximated by a smooth interpolation in the phase-field model, the width of which is controlled by $\Gamma_2$. As $\Gamma_2$ decreases, the phase-field simulations converge toward the sharp-interface solution (Figure \ref{fig:FDD}, inset).

\section{Three-Phase Results}\label{3 phase section}

To investigate the behaviour of our full three-phase system, we now consider the spontaneous formation of open gas cavities in a soft porous medium, starting from a nearly homogeneous initial distribution of gas within the pore space. This scenario mimics a natural system, such as a sea bed or lake bed, in which biological and/or chemical processes produce small gas bubbles throughout the otherwise undeformed and liquid-saturated pore space. As these bubbles grow, they may separate from the pore space by opening gas-rich (solid-free) cavities (Figure \ref{fig:Uniaxial}c). A similar situation could be achieved experimentally by saturating a porous packing with a fluid containing dissolved gas. Triggering gas exsolution would then lead to the formation and growth of gas bubbles. We study the evolution of this system through numerical simulations (\S \ref{Numerical Simulations}) and linear stability analysis (\S \ref{LSA section}).

\subsection{Numerical Simulations} \label{Numerical Simulations}

To focus on a porous medium that is sufficiently large that boundary effects are negligible in the bulk of the system, we impose periodic boundary conditions. We also set $Q_0=0$ without loss of generality; a non-zero value of $Q_0$ corresponds to bulk translation at a fixed rate. For initial conditions, we set
\begin{align}
    \phi_g(x,0) &= S_0\phi_0 +  \eta(x),\\
    \phi(x,0)&=\phi_0-|\phi^*|,
\end{align}
where $S_0$ is the initial gas saturation, $\eta$ is a field of small, random fluctuations, and $\phi^*\ll 1$ ensures an initial state of slight compression, so that the material is initially undamaged. 

\begin{figure}
    \centering
    \includegraphics{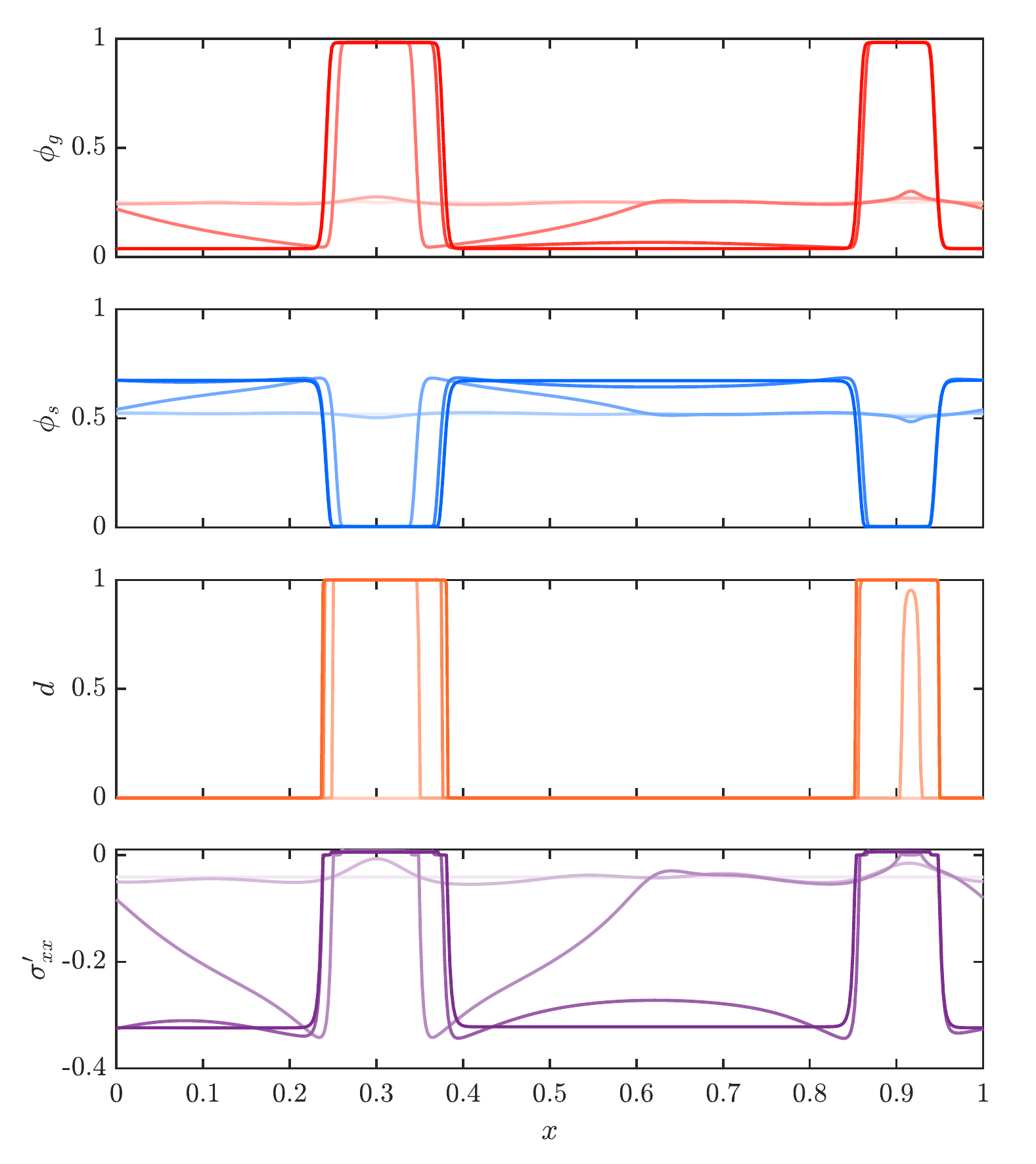}
    \caption{An initially homogeneous mixture will spontaneously separate into gas-rich cavities surrounded by a compressed gas-poor porous domain. Numerical simulations show the formation of open gas cavities as a result of phase separation. We show the distribution of $\phi_g$, $\phi_s$, $d$ and $\sigma'_{xx}$ at times $t=0$, $0.004$, $0.006$, $0.008$ and $1$ (light to dark colours). We use parameter values of $\phi_0=0.5$, $\phi^*=-0.02$, $S_0=0.55$, $\Theta=10^{-4}$, $\mathcal{M}_r=50$, $\mathcal{D}=2$, $\varsigma=0.1$, $\chi_{gl}=3.5$, $\chi_{gs}=3$, $\Gamma_i=10^{-4}$, $\kappa_c=10^{-4}$ and $\Lambda_d=10^{-8}$.}
    \label{fig:Cavity Formation}
\end{figure}

Evolving the system from this initial state, we see that perturbations in the gas fraction grow over time, deforming the solid skeleton (Figure \ref{fig:Cavity Formation}).
As gas bubbles outgrow the available pore space, they expand to form cavities in the solid by displacing solid grains. As the cavities form and then grow, the rest of the solid skeleton is increasingly compressed into a smaller region. A local equilibrium state is reached once elastic stresses balance the thermodynamic forcing of the phase separation. In this equilibrium state, the compressed solid packing has a uniform volume fraction throughout the porous medium. The formation of cavities damages the solid skeleton such that there is negligible elastic stress within the cavities.

\begin{figure}
    \centering
    \includegraphics{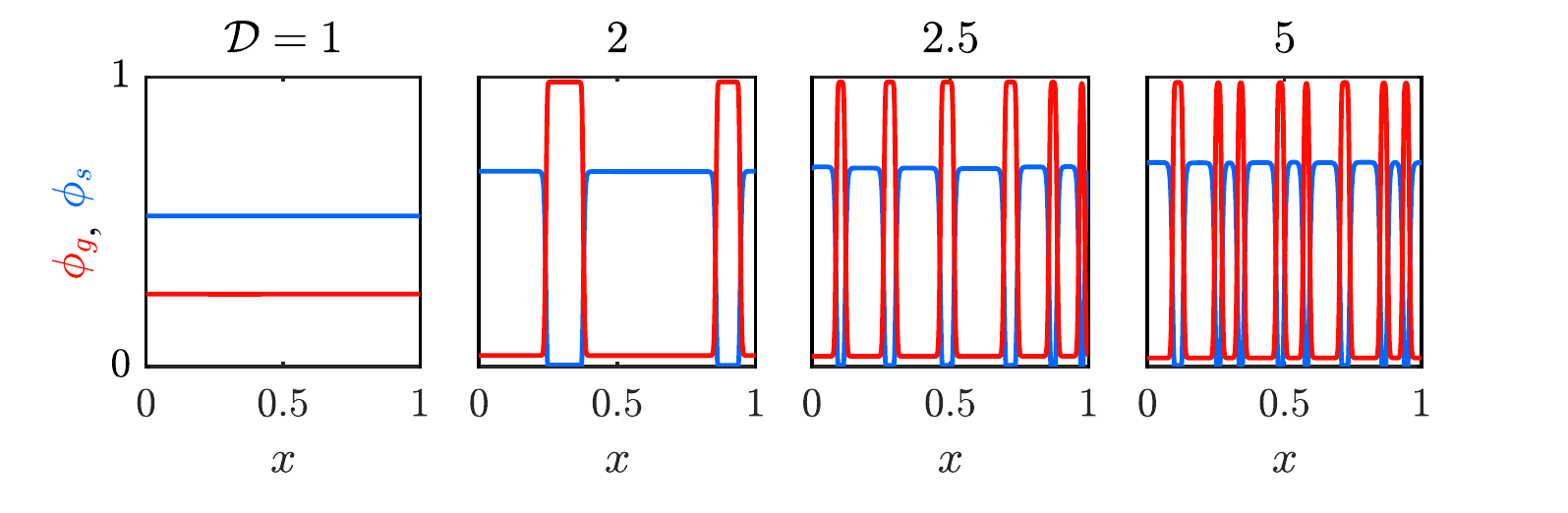}
    \caption{Steady-state distribution of $\phi_g$ (red) and $\phi_s$ (blue) from numerical simulations for a range of different characteristic deformabilities $\mathcal{D}$. Cavity formation is suppressed within a stiffer porous medium (low deformability), whereas a softer porous medium (high deformability) allows the formation of several macroscopic cavities. Other parameters are the same as in Figure \ref{fig:Cavity Formation}.}
    \label{fig:Deformability plot}
\end{figure}

We investigate the impact of the deformability of the porous medium on phase separation by running numerical simulations over a range of values of $\mathcal{D}$ (Figure~\ref{fig:Deformability plot}).
For a stiff porous medium ($\mathcal{D}\lesssim1$), the elastic forces within the solid skeleton are too strong for the gas to deform the packing. The gas thus remains within the pore space and phase separation is suppressed. Conversely, for a sufficiently soft porous medium ($\mathcal{D}\gtrsim1$), gas is able to open macroscopic cavities within the packing as described above. We investigate the impact of other parameters (such as $\chi_{gs}$ and $S_0$) in \S \ref{LSA section} below. 

\subsection{Linear Stability Analysis}\label{LSA section}

To identify the parameters that control the onset of phase separation, we now perform a linear stability analysis for our model. We linearise the system about a base state that represents an undamaged, precompressed porous matrix of porosity $\phi_0'=\phi_0~-~|\phi^*|$, whose pore space is occupied by a homogeneous distribution of gas with saturation $S_0$ and volume fraction $\phi_{g0}=S_0\phi_0$. We assume that perturbations of size $\epsilon\ll 1$ about this base state take the form of normal modes with growth rate $s$ and wavenumber $k$, such that the perturbed gas fraction and porosity are given by,
\begin{subequations}\label{LSA base states}
\begin{align}
    \phi_g&=\phi_{g0}+\epsilon \widehat{\phi}_g ~\mathrm{exp}\left(s t+ikx\right)+\mathcal{O}\left(\epsilon^2\right),\\
    \phi&=\phi_{0}'+\epsilon \widehat{\phi} ~\mathrm{exp}\left(s t+ikx\right)+\mathcal{O}\left(\epsilon^2\right),
\end{align}
\end{subequations}
where $\widehat{\phi}_g$ and $\widehat{\phi}$ characterise the $\mathcal{O}(\epsilon)$ contributions to $\phi_g$ and $\phi$, respectively.

Substituting this ansatz (Equations \ref{LSA base states}) into our model (Equations \ref{PhiGND}--\ref{KortewegND}) and linearising in $\epsilon$ leads to a set of equations that describe the evolution of small perturbations. To $\mathcal{O}(\epsilon)$, our model thus reduces to the following algebraic equations for $\widehat{\phi}_g$ and $\widehat{\phi}$,
\begin{subequations}\label{Linearised equations}
\begin{multline}
    s \widehat{\phi}_g=-\mathcal{M}_r\left(1-\phi_{g0}\right)\phi_{g0}\left[\left(h_{gg}k^2+\nu_{gg}k^4\right)\widehat{\phi}_g+\left(h_{gf}k^2+\nu_{gf}k^4\right)\widehat{\phi}\right]\\ +\left(\phi'_0-\phi_{g0}\right)\phi_{g0}\left[\left(h_{lg}k^2+\nu_{lg}k^4\right)\widehat{\phi}_g+\left(h_{lf}k^2+\nu_{lf}k^4\right)\widehat{\phi}\right],
\end{multline}
\begin{multline}
    s \widehat{\phi}=-\mathcal{M}_r\left(1-\phi'_{0}\right)\phi_{g0}\left[\left(h_{gg}k^2+\nu_{gg}k^4\right)\widehat{\phi}_g+\left(h_{gf}k^2+\nu_{gf}k^4\right)\widehat{\phi}\right]\\ -\left(1-\phi'_0\right)\left(\phi'_0-\phi_{g0}\right)\left[\left(h_{lg}k^2+\nu_{lg}k^4\right)\widehat{\phi}_g+\left(h_{lf}k^2+\nu_{lf}k^4\right)\widehat{\phi}\right].
\end{multline}
\end{subequations}
In the above, we have introduced the functions $h_{\alpha\beta}=\pd{h_\alpha}{\phi_\beta}|_{\phi'_0,\phi_{g0}}$, where $h_\alpha$ is the homogeneous part of the chemical potential of phase $\alpha$, and the parameters $\nu_{\alpha\beta}$, which are different combinations of interfacial coefficients. In general, $h_{\alpha\beta}$ and $\nu_{\alpha\beta}$ depend on the dimensionless groups defined in \S \ref{Uniaxial} and the base states $\phi_{g0}$ and $\phi'_0$, but they are constant with respect to $k$ and $s$. Explicit expressions for $h_{\alpha\beta}$ and $\nu_{\alpha\beta}$ are presented in Appendix \ref{LSA Details}. Collecting like terms and finding the eigenvalues of Equations (\ref{Linearised equations}) leads to the dispersion relation
\begin{equation}\label{Dispersion}
    s^2+\zeta_1\left(k\right)k^2s+\zeta_2\left(k\right)k^4=0,
\end{equation}
where $\zeta_1\left(k\right)=a_1+a_2k^2$ and $\zeta_2\left(k\right)=b_1+b_2k^2+b_3k^4$. Explicit forms of $a_i$ and $b_i$ are given in Appendix \ref{LSA Details}.

As noted above, $s$ is the growth rate of small perturbations to the homogeneous system described by Equations (\ref{LSA base states}). The dispersion relation allows us to calculate $s$ as a function of the wavenumber, $k$, of a particular perturbation. If $s<0$ for all values of $k$, then perturbations will decay and the system is stable --- gas will remain in the pore space of the solid skeleton. If $s>0$ for any value of $k$, then perturbations with this particular wavenumber (or range of wavenumbers) will grow exponentially, and the system will be unstable --- phase separation will lead to macroscopic gas cavities. The solution of Equation (\ref{Dispersion}) is
\begin{equation}\label{growth}
   s = -\frac{\zeta_1k^2}{2} \pm \frac{k^2}{2}\sqrt{\zeta_1^2-4\zeta_2},
\end{equation}
with small-$k$ limit
\begin{equation}
    s \approx -\frac{a_1k^2}{2} \pm \frac{k^2}{2}\sqrt{a_1^2-4b_1},
\end{equation}
which shows that $s>0$ at small $k$ whenever either $a_1<0$ or $b_1<0$. In these cases, all perturbations with wavenumbers between zero and some cut-off value $k_c$ will be unstable and lead to phase separation. The cut-off value $k_c$ is defined by the non-zero roots of $\zeta_2\left(k\right)$. In addition to the region of instability for $k\in[0,k_c]$, our system has the unusual characteristic of forming an unstable band of wavenumbers $k\in[k_a,k_b]$, such that $k_a>0$ under certain circumstances. This characteristic occurs when $a_1>0$ and $b_1>0$, and $\zeta_2\left(k\right)$ has two distinct non-zero roots, which requires that both $b_2<0$ and $4b_1b_3<b_2^2$. The condition for stability is thus that
\begin{equation}\label{stability condition}
    a_1>0 \quad\textrm{and}\quad b_1>0\quad \textrm{and} \quad\left( b_2>0 ~~\textrm{or}~~ 4b_1b_3>b_2^2 \right).
\end{equation}
As $a_i,~b_i$ are complicated functions of many different parameters, we further explore the consequence of these stability conditions via a phase plane. Specifically, we plot the spinodal (or neutral stability) curves, where $s=0$, which separate stable and unstable regions of the parameter space. Equation (\ref{growth}) also reveals the possibility of oscillatory modes of instability, for which $s$ is complex, under certain parameter combinations. For the parameter space investigated below, these oscillatory modes occur so close to the spinodal curves that they are not observed in the simulations, and, as such, we do not explore them any further here.

\subsubsection{Onset of Phase Separation}

Three key parameters with regard to the formation of gas cavities are the deformability of the solid matrix, $\mathcal{D}$, the strength of the interaction between the gas and solid phases, $\chi_{gs}$, and the initial gas saturation, $S_0$. Recall that $\mathcal{D}$ measures the ratio of mixing energy to elastic energy, whereas $\chi_{gs}$ measures the energetic cost of gas--solid interactions. Typically, $\chi_{gs}$ is a function of physical properties such as the size and wetting characteristics of the solid particles. If the grains are smaller or if the gas phase is more strongly non-wetting to the solid, then the energetic cost of gas invading the pore space will be larger, resulting in a larger value of $\chi_{gs}$. The initial gas saturation $S_0$ measures how much gas is present in the system. If there is insufficient gas present, then capillary effects will not be strong enough to open macroscopic cavities and phase separation will not occur.

\begin{figure}
    \centering
    \includegraphics{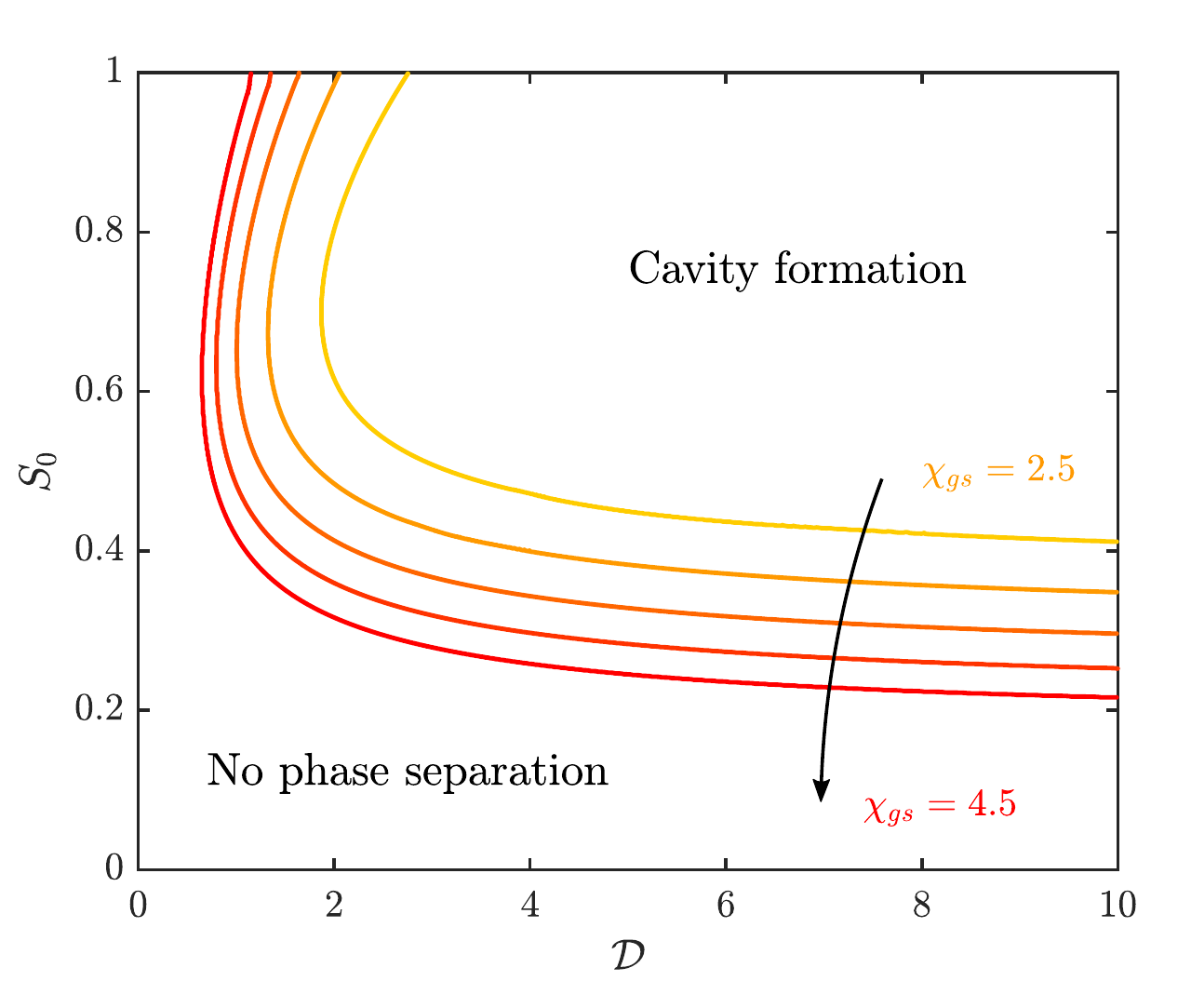}
    \caption{A phase plane showing the regions of stability (no phase separation) and instability (cavity formation) calculated from the linear stability analysis of the full 1D system. Spinodal curves are plotted for a range of values of the gas-solid interaction parameter, $\chi_{gs}$. An increase in $\chi_{gs}$ corresponds to an increase in capillary potential due to, for example, a smaller grain size, making it more energetically costly for gas to remain within the pore space and therefore promoting phase separation. All other parameters are the same as in Figure \ref{fig:Cavity Formation}.
    }
    \label{fig:Phase Plane}
\end{figure}

The stability condition defined in Equation (\ref{stability condition}) allows us to identify the regions of the parameter space in which phase separation occurs and the regions in which it does not. In Figure \ref{fig:Phase Plane}, we plot the spinodal curve for various values of $\chi_{gs}$ in the $\mathcal{D}-S_0$ plane. 
This phase plane shows the physically intuitive behaviour described above. For a particular value of $\chi_{gs}$, phase separation occurs in the top-right corner of the phase plane and is suppressed in the bottom-left corner. Phase separation is promoted as $\chi_{gs}$ increases, because if the interaction energy between the gas and solid phases is larger, then the energetic cost of their mixing is larger. If $S_0$ is too low, then phase separation will not occur for any particular value of $\chi_{gs}$.

The deformability of the solid skeleton has a strong influence on the onset of phase separation. At low $\mathcal{D}$, the solid matrix is too stiff to be deformed by capillary forces and the gas will remain within the pore space. For larger values of $\mathcal{D}$, however, the solid skeleton is more easily deformed, and capillary forces may be able to overcome the elastic resistance and open macroscopic cavities, provided that $S_0$ is large enough (\textit{i.e.,} that enough gas is present). Note that, for a particular value of $\chi_{gs}$, there is a critical value of $\mathcal{D}$ below which phase separation cannot occur, regardless of $S_0$.

\subsection{Characteristic Cavity Size}\label{Cavity Size}

\begin{figure}
    \centering
    \includegraphics[width=\textwidth]{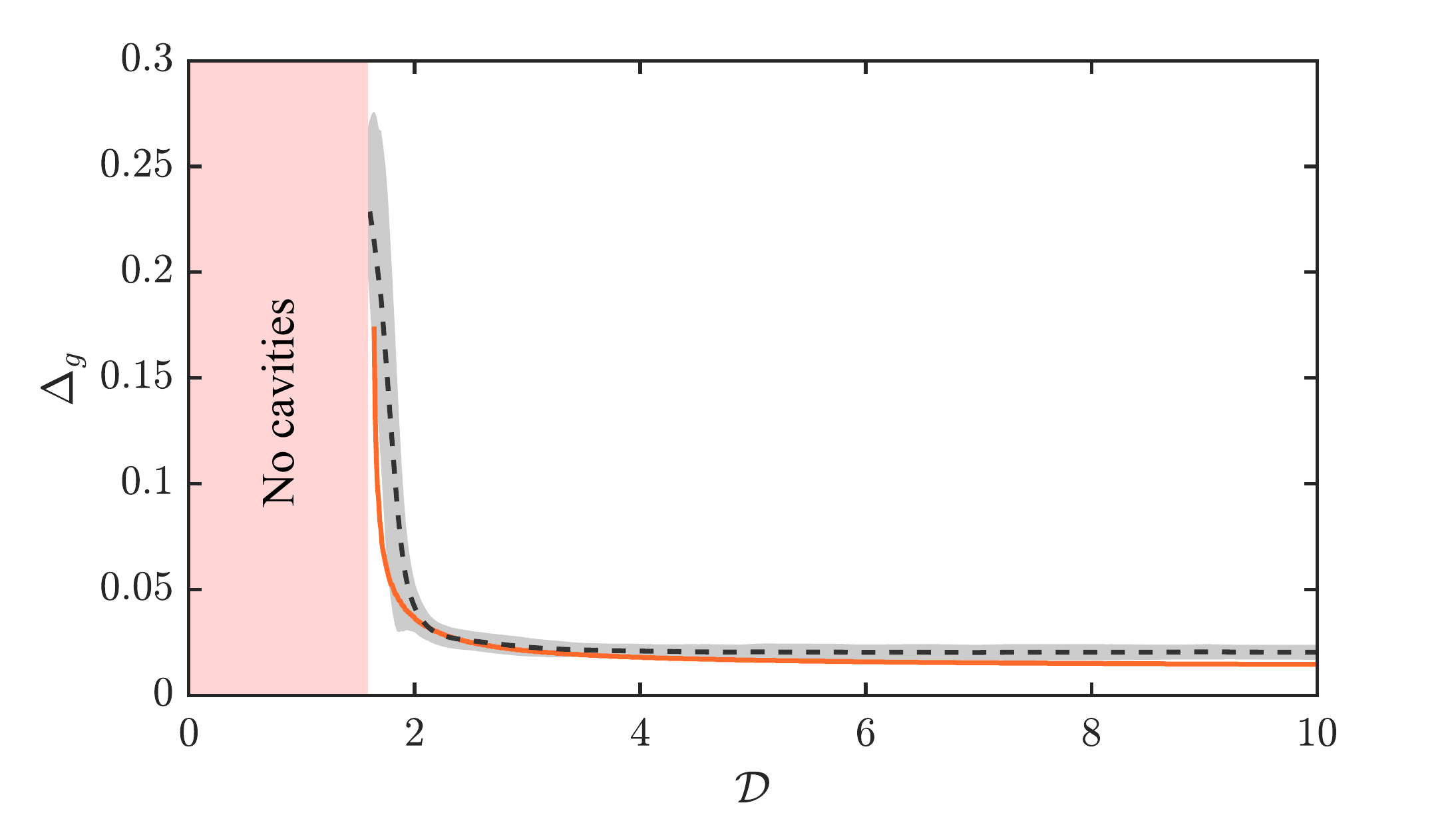}
    \caption{Characteristic size of gas cavities, $\Delta_g$, as a function of deformability, $\mathcal{D}$. from numerical simulations (black) and linear stability analysis (orange). The numerical results are the mean (black) and standard deviation (grey band) of 50 different initial conditions at each value of $\mathcal{D}$. All other parameters are the same as in Figure \ref{fig:Cavity Formation}.}
    \label{fig:Size plot}
\end{figure}

In addition to the parameters that control the onset of phase separation, we also investigate the characteristic size of the gas cavities, $\Delta_g$. The characteristic cavity size can be estimated from the linear stability analysis by finding the wavenumber of the most unstable mode at each point in the parameter space, $k^*$, and estimating $\Delta_g\approx\frac{\pi}{k^*}$. Doing so for a range of $\mathcal{D}$ then allows us to predict cavity size as a function of deformability. We compare this prediction with the mean cavity size at steady state from numerical simulations (Figure~\ref{fig:Size plot}).

As $\mathcal{D}$ increases, $\Delta_g$ decreases. However, the total amount of gas within the cavities increases with $\mathcal{D}$, so this decrease in $\Delta_g$ is more than offset by an increase in the total number of cavities. This behaviour can also be seen qualitatively in Figure \ref{fig:Deformability plot}: there are more, smaller cavities  for $\mathcal{D}=5$ than for $\mathcal{D}=2$ and $\mathcal{D}=2.5$. Intuitively, we would expect larger cavities to form in a softer material since the solid skeleton can undergo larger deformations. Indeed, the total `cavity volume' is larger for softer materials, but these cavities are smaller and more numerous. We attribute this observation to the fact that, in a softer material, a smaller amount of gas is needed locally to open a cavity. For a stiff porous material, a large amount of gas will need to accumulate in order to force open a cavity, and so when the cavity is formed, it will be correspondingly larger.

Our numerical simulations show remarkably good qualitative and quantitative agreement with the linear stability analysis for the value of $\Delta_g$, which is surprising given the strongly nonlinear nature of this process. Figure \ref{fig:Size plot} also shows that, as expected, there exists a certain critical deformability below which phase separation does not occur. The linear stability analysis and numerical simulations show this transition occurring at around the same value of $\mathcal{D}$.

\section{Conclusions} \label{Conclusions section}

The key feature of two-phase flow in a deformable porous medium relative to a rigid porous medium is the ability of the non-wetting phase to form open cavities within the solid skeleton. Motivated by the formation of non-wetting gas cavities in sediments, we have derived a thermodynamically consistent phase-field model that describes the formation of cavities within a soft porous material as a result of gas-liquid-solid phase separation, and have explored our model in a one-dimensional setting. In the two limiting cases of no-solid and no-gas, our model reproduces the expected behaviour for these well-studied systems. When no solid is present, we reproduce the fluid-fluid phase separation and coarsening behaviour characteristic of the Cahn-Hilliard equation; when no gas is present, our model matches well with analytic solutions to traditional sharp-interface Biot poroelasticity.

In the full three-phase system, phase separation is inhibited by elastic resistance from the solid, which imposes limits on the conditions under which cavities form: if the solid skeleton is too stiff, gas will remain within the pore space. We have shown how the onset of phase separation depends on different parameters, including the deformability of the solid matrix and the wetting characteristics of the two fluids, via both linear stability analysis and numerical simulation. We have also shown that, for a softer porous material, we expect smaller, but more numerous, cavities than for a stiffer porous material, owing to a smaller amount of gas needing to accumulate within the pore space in order to open a cavity in the former case.

{Our work has been motivated by the venting of gas from non-cohesive granular sediments, in which the compressibility of the gas phase can be neglected when the capillary entry pressure of the granular skeleton is sufficiently small compared to the ambient liquid pressure. Our model is also relevant for different fluid pairs in other contexts, such as phase-separating liquid droplets in polymer networks \cite{Style2018,Rosowski2020,FernandezRico2022}. In polymer systems, the porous skeleton does not have a granular microstructure, and the energy required to fracture the polymer network to form cavities affects the final size of droplets \cite{Vidal-Henriquez2021,Ronceray2022,Kim2020}. Our model can capture fluid-fluid phase separation within such porous materials through an appropriate choice of the Griffith fracture energy.}

We have solved our model assuming a neo-Hookean elastic response for the solid skeleton, but our theory allows for other elastic laws. We have assumed that the gas phase is incompressible, and that the gas and liquid phases are immiscible. We anticipate that relaxing these assumptions will lead to further interesting physical phenomena, the investigation of which will be the subject of future work. Our model makes use of a phenomenological free energy. Connecting the parameters used in our free energy to physical quantities remains an avenue for future study; considering the pore-scale thermodynamics of a gas-liquid-solid mixture could provide this link. Ongoing experimental work will allow us to compare our theory with experimental observations, and will focus on the dynamic transition between the phase-separated and homogeneous regimes via the application of external confining stress.

\begin{acknowledgments}
This research was supported by the European Research Council (ERC) under the European Union's Horizon 2020 Programme [Grant No. 805469], and by the Engineering and Physical Sciences Research Council (EPSRC) [Grant No. EP/S034587/1]. The authors also acknowledge the use of the University of Oxford Advanced Research Computing (ARC) facility [http://dx.doi.org/10.5281/zenodo.22558].
\end{acknowledgments}

\appendix

\section{Solid Mechanics}\label{Detailed Mechanics}

As noted in \S \ref{Free Energy}, our system displays distinct mechanical behaviours depending on whether the material is in tension or compression. As such, we must identify when each regime occurs. Following the approach of Tang \textit{et al.} \cite{Tang2019}, we first rewrite the neo-Hookean strain-energy density in terms of the principal stretches, $\lambda_i$,
\begin{equation}
    \mathcal{W}_{el} = \frac{1}{2}\mathcal{G}\sum_{i=1}^3\left(\lambda_i^2-1-2\log{\lambda_i}\right)+\frac{1}{2}\mathcal{K}\left(\log{J}\right)^2,
\end{equation}
where we have used that $J=\lambda_1\lambda_2\lambda_3$. The first part of this expression corresponds to stretching deformations and the second part to volumetric deformations. We then decompose this strain-energy function into tensile and compressive parts, given by $\mathcal{W}^+=\mathcal{W}_{el}\left(\lambda^+_i,J^+\right)$ and $\mathcal{W}^-=\mathcal{W}_{el}\left(\lambda^-_i,J^-\right)$, respectively. We define $\lambda_i^{\pm}$ and $J^{\pm}$ as piece-wise functions, with
\begin{equation}
    \lambda_i^+ = \begin{cases}
    1 &\textrm{for}~\lambda_i\leq 1,\\
    \lambda_i & \textrm{for}~\lambda_i>1,
    \end{cases}\qquad
    J^+ = \begin{cases}
    1 &\textrm{for}~J\leq 1,\\
    J & \textrm{for}~J>1
    \end{cases}
\end{equation}
and
\begin{equation}
    \lambda_i^- = \begin{cases}
    \lambda_i & \textrm{for}~\lambda_i<1,\\
    1 &\textrm{for}~\lambda_i\geq 1,
    \end{cases}\qquad
    J^- = \begin{cases}
    J & \textrm{for}~J<1,\\
    1 &\textrm{for}~J\geq 1.
    \end{cases}
\end{equation}
As per Equations (\ref{Psi2}), the tensile part of the elastic energy, $\mathcal{W}^+$ is degraded by an increase in the damage parameter.

The effective stress, $\boldsymbol{\sigma}'$, is found by taking the derivative of the elastic free energy (Equation \ref{Stress definition}),
\begin{equation}\label{appendix3dstress}
    \boldsymbol{\sigma}'=g\left(d\right)\frac{1}{J}\frac{\partial\mathcal{W}^+}{\partial\mathbb{F}}\mathbb{F}^{\intercal}+\frac{1}{J}\frac{\partial\mathcal{W}^-}{\partial\mathbb{F}}\mathbb{F}^{\intercal}.
\end{equation}
Using the chain rule, we have that (following \cite{Tang2019}),
\begin{equation}\label{dW+-Df}
    \pd{\mathcal{W}^+}{F_{ij}}=\sum_{k=1}^3 \pd{\mathcal{W}^+}{\lambda_k^+}\pd{\lambda_k^+}{F_{ij}} +\pd{\mathcal{W}^+}{J^+}\pd{J^+}{F_{ij}}, \quad
    \pd{\mathcal{W}^-}{F_{ij}}=\sum_{k=1}^3 \pd{\mathcal{W}^-}{\lambda_k^-}\pd{\lambda_k^-}{F_{ij}} +\pd{\mathcal{W}^-}{J^-}\pd{J^-}{F_{ij}},
\end{equation}
where
\begin{subequations}
\begin{equation}
    \pd{\lambda_k^+}{F_{ij}} = \begin{cases}
    0 &\textrm{for}~\lambda_k\leq1,\\
    \pd{\lambda_k}{F_{ij}} & \textrm{for}~\lambda_k>1,
    \end{cases} \qquad
    \pd{\lambda_k^-}{F_{ij}} = \begin{cases}
    \pd{\lambda_k}{F_{ij}} & \textrm{for}~\lambda_k<1,\\
    0 &\textrm{for}~\lambda_k\geq1,
    \end{cases}
\end{equation}
\begin{equation}
    \pd{J^+}{F_{ij}} = \begin{cases}
    0 &\textrm{for}~J\leq1,\\
    \pd{J}{F_{ij}} & \textrm{for}~J>1,
    \end{cases} \qquad
    \pd{J^-}{F_{ij}} = \begin{cases}
    \pd{J}{F_{ij}} & \textrm{for}~J<1,\\
    0 &\textrm{for}~J\geq1
    \end{cases}
\end{equation}
\end{subequations}
and
\begin{subequations}\label{dW+-dlambda}
\begin{equation}
    \pd{\mathcal{W}^+}{\lambda_k^+}=\mathcal{G}\left(\lambda_k^+-\frac{1}{\lambda_k^+}\right), \qquad
    \pd{\mathcal{W}^-}{\lambda_k^-}=\mathcal{G}\left(\lambda_k^--\frac{1}{\lambda_k^-}\right),
\end{equation}
\begin{equation}
    \pd{\mathcal{W}^+}{J^+}=\mathcal{K}\frac{\log{J}^+}{J^+}, \qquad
    \pd{\mathcal{W}^-}{J^-}=\mathcal{K}\frac{\log{J}^-}{J^-}.
\end{equation}
\end{subequations}
This constitutive law reduces to a standard neo-Hookean behaviour in the case of an undamaged system ($d=0$).

The solid mechanics of our system greatly simplify in the uniaxial case, as described in \S \ref{Uniaxial}. In 1D, $\mathbf{u}_s = u_s\left(x,t\right) \mathbf{\hat{e}}_x$, and hence using Equation (\ref{F in u}), we see that the deformation gradient tensor is diagonal, with
\begin{equation}
\fixTABwidth{T}
\mathbb{F}^{-1} = 
\parenMatrixstack{
1 - \frac{\partial u_s}{\partial x} & 0 & 0\\
0 & 1 & 0\\
0 & 0 & 1},
\end{equation}
Noting that $J=\textrm{det}\left(\mathbb{F}\right)$, we thus write that
\begin{equation}
\fixTABwidth{T}
\mathbb{F} = 
\parenMatrixstack{
~J~ & 0 & 0\\
0 & 1 & 0\\
0 & 0 & 1}.
\end{equation}
In this case, the principal stretches $\lambda_i$ are the eigenvalues of the deformation gradient, and so only one of these stretches $\lambda=\lambda_1=J$ is non-constant in this system. As such, the neo-Hookean strain-energy can be written solely in terms of $J$, and we can identify tension as being when $J>1$. The above constitutive behaviour (Equation \ref{appendix3dstress}) thus simplifies to
\begin{equation}
    \sigma'_{xx}=g\left(d\right)\pd{\mathcal{W}^+}{J}+\pd{\mathcal{W}^-}{J},
\end{equation}
with the tensile and compressive components of the free energy defined as,
\begin{equation}
    \mathcal{W}^+= \begin{cases}
    0 &\textrm{for}~J\leq 1,\\
    \mathcal{W}_{el}^0 & \textrm{for}~J>1,
    \end{cases}
\end{equation}
and
\begin{equation}
    \mathcal{W}^-= \begin{cases}
    \mathcal{W}_{el}^0 & \textrm{for}~J<1,\\
    0 &\textrm{for}~J\geq 1,
    \end{cases}
\end{equation}
respectively, where $\mathcal{W}_{el}^0=\frac{1}{2}\mathcal{G}\left(J^2-1-2\log{J}\right)+\frac{1}{2}\mathcal{K}\left(\log{J}\right)^2$. Undertaking the differentiation of the free energy with respect to $J$ then gives,
\begin{equation}
    \sigma'_{xx}=\begin{cases}
    \sigma'_0\left(\phi\right) &J\leq1,\\
    \sigma'_0\left(\phi\right)   g\left(d\right) &J>1,
    \end{cases}
\end{equation}
where $\sigma'_0$ is the undamaged neo-Hookean stress, given by,
\begin{equation}
    \sigma'_0=\mathcal{G}\left(J-\frac{1}{J}\right)+\mathcal{K}\frac{\log{J}}{J}.
\end{equation}
To complete our description of the mechanics, we link $J$ to the porosity, $\phi$, using Equation (\ref{Incompressibility}).

\section{Capillary Potential}\label{Capillary Potential}

To formulate explicit expressions for the capillary potentials, we start by substituting the free energy of the system into Equation (\ref{Capillary definition}). The liquid and gas potentials contain three distinct parts,
\begin{equation}
    \Psi_{\alpha} = \frac{\partial\psi_{\mathrm{bulk}}}{\partial\varphi_\alpha} +\frac{\partial\psi_{\mathrm{mix}}}{\partial\varphi_\alpha}- \nabla_0\cdot\left(\frac{\partial\psi_{\mathrm{interface}}}{\partial\nabla_0\varphi_\alpha}\right),
\end{equation}
each associated with different components of the free energy. The free energy is written in \S \ref{Free Energy} as a function of Eulerian variables, $\phi_\alpha$ and $\nabla$, whereas the capillary potential is defined in terms of derivatives with respect to Lagrangian variables, $\varphi_\alpha$ and $\nabla_0$. We thus use the relations $\varphi_\alpha=\phi_\alpha J$, $\nabla=\mathbb{F}^{-\intercal}\nabla_0$, and $J=\varphi_s^0+\varphi_g+\varphi_l$ to give
\begin{subequations}
\begin{equation}
    \Psi_{g} = \rho_g\mu^0_g +\Pi_g -\gamma_1 \nabla^2\phi_g
    -\frac{\gamma_3}{2}\nabla^2\phi_l,
\end{equation}
\begin{equation}
    \Psi_{l} = \rho_l\mu^0_l +\Pi_l -\gamma_2\nabla^2 \phi_l
    -\frac{\gamma_3}{2}\nabla^2\phi_g,
\end{equation}
\end{subequations}
where we have further defined the interaction potentials, $\Pi_\alpha$, as the component of the capillary potential resulting from the differentiation of $\psi_{\mathrm{mix}}$. These are given by:
\begin{subequations}
\begin{equation}\label{PiG}
    \Pi_g = \mathcal{E}_{\mathrm{mix}}\left[
    \log{\phi_g} + \phi_s + \left(1-\phi_g\right)\left(\chi_{gl}\phi_l + \chi_{gs}\phi_s\right) \right],
\end{equation}
\begin{equation}\label{PiL}
    \Pi_l = \mathcal{E}_{\mathrm{mix}}\left[\log{\phi_l} + \phi_s + \phi_g\left(\chi_{gl}\phi_g - \chi_{gs}\phi_s\right)\right].
\end{equation}
\end{subequations}
For incompressible, immiscible fluids, $\rho_\alpha\mu_\alpha^0$ is constant. In our model, the capillary potential only appears as the argument of a gradient function, and so this constant bulk term has no effect on the overall state of the system. The spatial derivatives in the capillary potentials resist the formation of sharp gradients in volume fraction, and serve to regularise the system by preventing the formation of shocks in the gas fraction and porosity.

\section{Korteweg Stress} \label{Korteweg Stress}

In order to derive an explicit expression for the Korteweg stress, we first consider the free energy of a general interface between two unspecified phases, $\alpha$ and $\beta$, using the form given in Equation (\ref{psi int general}),
\begin{equation}
    \psi_{\alpha\beta}=\frac{\gamma}{2}\left(\nabla\phi_\alpha\cdot\nabla\phi_\beta\right) J.
\end{equation}
As derived in \S \ref{Constitutive Behaviour}, the Korteweg stress has two distinct contributions, which we denote $\mathbb{K}_1$ and $\mathbb{K}_2$, where,
\begin{equation}
    \mathbb{K}_1=\frac{1}{J}\frac{\partial\psi_{\mathrm{interface}}}{\partial\mathbb{F}}\mathbb{F}^{\intercal}, \qquad \mathbb{K}_2=- \left(\nabla_0\cdot\frac{\partial\psi_{\mathrm{interface}}}{\partial\nabla_0 J}\right)\mathbb{I}.
\end{equation}
In order to carry out this differentiation, we need to write $\psi_{\alpha\beta}$ in terms of Lagrangian quantities. Using the conversions $\varphi_\alpha=\phi_\alpha J$ and $\nabla=\mathbb{F}^{-\intercal}\nabla_0$, we rewrite the interfacial free energy as,
\begin{equation}\label{interfacial energy expanded}
    \psi_{\alpha\beta}=\frac{\gamma}{2J^3}\left[J\nabla_0\varphi_\alpha-\varphi_\alpha\nabla_0J\right]\cdot\mathbb{C}^{-1}\cdot\left[J\nabla_0\varphi_\beta-\varphi_\beta\nabla_0J\right],
\end{equation}
where $\mathbb{C}=\mathbb{F}^\intercal\mathbb{F}$ is the right Cauchy-Green tensor. We can then differentiate Equation (\ref{interfacial energy expanded}) to find the Korteweg stress. Note that we carry out the derivative of $J$ with respect to $\mathbb{F}$ using Jacobi's formula, $\pd{J}{\mathbb{F}}=J\mathbb{F}^{-\intercal}$. The result is
\begin{equation}
    \mathbb{K}_1=-\frac{\gamma}{2}\left[\nabla\phi_\alpha\otimes\nabla\phi_\beta+\nabla\phi_\beta\otimes\nabla\phi_\alpha+\left(\nabla\phi_\alpha\cdot\nabla\phi_\beta\right) \mathbb{I}\right]
\end{equation}
and
\begin{equation}
    \mathbb{K}_2=\frac{\gamma}{2}\left(2\nabla\phi_\alpha\cdot\nabla\phi_\beta+\phi_\alpha\nabla^2\phi_\beta+\phi_\beta\nabla^2\phi_\alpha\right)\mathbb{I},
\end{equation}
where $\otimes$ is the tensor product, and we have converted back to Eulerian variables. Combining these two expressions gives the total Korteweg stress resulting from an $\alpha-\beta$ interface,
\begin{equation}
    \mathbb{K}_{\alpha\beta}=\frac{\gamma}{2}\left( \nabla\phi_\alpha\cdot\nabla\phi_\beta + \phi_\alpha\nabla^2\phi_\beta + \phi_\beta\nabla^2\phi_\alpha \right) \mathbb{I} 
- \frac{\gamma}{2} \left(\nabla\phi_\alpha\otimes\nabla\phi_\beta + \nabla\phi_\beta\otimes\nabla\phi_\alpha\right).
\end{equation}
The three types of interface in our system are gas-gas, liquid-liquid and gas-liquid. The Korteweg stress is then the sum of the contributions from these interfaces,
\begin{multline}\label{Korteweg}
\mathbb{K} = \gamma_1\left( \frac{1}{2}\vert\nabla\phi_g\vert^2 + \phi_g\nabla^2\phi_g \right) \mathbb{I} 
- \gamma_1 \nabla\phi_g\otimes\nabla\phi_g 
+ \gamma_2\left( \frac{1}{2}\vert\nabla\phi_l\vert^2 + \phi_l\nabla^2\phi_l \right) \mathbb{I} 
- \gamma_2 \nabla\phi_l\otimes\nabla\phi_l \\
+ \frac{\gamma_3}{2}\left( \nabla\phi_g\cdot\nabla\phi_l + \phi_g\nabla^2\phi_l + \phi_l\nabla^2\phi_g \right) \mathbb{I} 
- \frac{\gamma_3}{2} \left(\nabla\phi_g\otimes\nabla\phi_l + \nabla\phi_l\otimes\nabla\phi_g\right).
\end{multline}
In 1D, only the $xx$-component of $\mathbb{K}$ is relevant, and is given by
\begin{equation}
    \begin{split}
        K_{xx} = 
\gamma_1\left[\phi_g\frac{\partial^2\phi_g}{\partial x^2}-\frac{1}{2}\left(\frac{\partial\phi_g}{\partial x}\right)^2\right]
&+\gamma_2\left[\phi_l\frac{\partial^2\phi_l}{\partial x^2}-\frac{1}{2}\left(\frac{\partial\phi_l}{\partial x}\right)^2\right]\\
&+\frac{1}{2}\gamma_3\left[\phi_g\frac{\partial^2\phi_l}{\partial x^2}+\phi_l\frac{\partial^2\phi_g}{\partial x^2}-\frac{\partial\phi_l}{\partial x}\frac{\partial\phi_g}{\partial x}\right].
    \end{split}
\end{equation}
This expression for the Korteweg stress is used in the uniaxial system described in \S \ref{Uniaxial}.

\section{Model Summary}\label{Full Model Appendix}

In this appendix we present a summary of the governing equations derived in our model. We note that gas, liquid, and solid volume fractions are related by $\phi_g+\phi_l+\phi_s=1$. All quantities in this appendix are dimensional. Conservation of mass can be written
\begin{subequations}
\begin{align}
    \frac{\partial\phi_g}{\partial t} + \nabla\cdot\left(\phi_g\mathbf{q}\right) +\nabla\cdot\left[ \left(1 - \phi_g\right) \mathbf{w}_g - \phi_g\mathbf{w}_l \right]&=0,\\
    \frac{\partial\phi}{\partial t} + \nabla\cdot\left(\phi\mathbf{q}\right) +\nabla\cdot\left[ \left( 1 -\phi\right) \left(\mathbf{w}_g + \mathbf{w}_l\right)\right]&=0,\\
    \nabla\cdot\mathbf{q} &= 0,
\end{align}
\end{subequations}
where $\phi=\phi_g+\phi_l$ is the porosity. The relative fluid fluxes are given by
\begin{equation}
      \mathbf{w}_{\alpha}=-\phi_\alpha \mathcal{M}_\alpha^0\nabla\left(p+\Psi_{\alpha}\right) ,
\end{equation}
where pressure gradients are balanced by a sum of elastic and Korteweg stresses,
\begin{equation}
      \nabla p = \nabla\cdot\left(\boldsymbol{\sigma}'+\mathbb{K}\right).
\end{equation}
For our specific choice of free energy, the capillary potentials are
\begin{subequations}
\begin{equation}
    \Psi_{g} = \rho_g\mu^0_g +\Pi_g -\gamma_1 \nabla^2\phi_g
    -\frac{\gamma_3}{2}\nabla^2\phi_l,
\end{equation}
\begin{equation}
    \Psi_{l} = \rho_l\mu^0_l +\Pi_l -\gamma_2\nabla^2 \phi_l
    -\frac{\gamma_3}{2}\nabla^2\phi_g,
\end{equation}
\end{subequations}
where the interaction potentials are
\begin{subequations}
\begin{equation}
    \Pi_g = \mathcal{E}_{\mathrm{mix}}\left[
    \log{\phi_g} + \phi_s + \left(1-\phi_g\right)\left(\chi_{gl}\phi_l + \chi_{gs}\phi_s\right) \right],
\end{equation}
\begin{equation}
    \Pi_l = \mathcal{E}_{\mathrm{mix}}\left[\log{\phi_l} + \phi_s + \phi_g\left(\chi_{gl}\phi_g - \chi_{gs}\phi_s\right)\right].
\end{equation}
\end{subequations}
The effective stress is given by
\begin{equation}
    \boldsymbol{\sigma}'=g\left(d\right)\frac{1}{J}\frac{\partial\mathcal{W}^+}{\partial\mathbb{F}}\mathbb{F}^{\intercal}+\frac{1}{J}\frac{\partial\mathcal{W}^-}{\partial\mathbb{F}}\mathbb{F}^{\intercal},
\end{equation}
with $g\left(d\right)=\left[\left(1-\Theta\right)\left(1-d\right)^2+\Theta\right]$ and where $\frac{\partial\mathcal{W}^\pm}{\partial\mathbb{F}}$ for a neo-Hookean solid skeleton given in Equations (\ref{dW+-Df})-(\ref{dW+-dlambda}). The Korteweg stress is given by
\begin{multline}
\mathbb{K} = \gamma_1\left( \frac{1}{2}\vert\nabla\phi_g\vert^2 + \phi_g\nabla^2\phi_g \right) \mathbb{I} 
- \gamma_1 \nabla\phi_g\otimes\nabla\phi_g 
+ \gamma_2\left( \frac{1}{2}\vert\nabla\phi_l\vert^2 + \phi_l\nabla^2\phi_l \right) \mathbb{I} 
- \gamma_2 \nabla\phi_l\otimes\nabla\phi_l \\
+ \frac{\gamma_3}{2}\left( \nabla\phi_g\cdot\nabla\phi_l + \phi_g\nabla^2\phi_l + \phi_l\nabla^2\phi_g \right) \mathbb{I} 
- \frac{\gamma_3}{2} \left(\nabla\phi_g\otimes\nabla\phi_l + \nabla\phi_l\otimes\nabla\phi_g\right).
\end{multline}
The damage parameter, $d$, evolves according to
\begin{equation}
    \left[4l_d\left(1-\Theta^{-1}\right)\mathcal{W}^++\mathcal{G}_c\right]\left(1-d\right)+4\mathcal{G}_cl_d^2\nabla_0^2d=\mathcal{G}_c,
\end{equation}
where $\mathcal{W}^\pm$ are defined in Appendix \ref{Detailed Mechanics}. Finally, we link deformation to porosity via $\mathbb{F} = \left(\mathbb{I}-\nabla\mathbf{u}_s\right)^{-1}$ and $J=\textrm{det}(\mathbb{F})=\frac{1-\phi_0}{1-\phi}$.

\section{Analytical Solution to Sharp-Interface Poroelasticity}\label{Analytic}

To find a steady-state solution to the sharp-interface model described in \S \ref{LiquidSolidSection}, we first note that the porosity is described by a piecewise function,
\begin{equation}
    \phi\left(x\right)=\begin{cases}
    1 & x < \delta,\\
    f\left(x\right) & \delta \leq x \leq 1.
        \end{cases}
\end{equation}
To find an analytical solution for $f\left(x\right)$, we solve Equation (\ref{SI poroelasticity}) over the domain $x\in\left[\delta\left(t\right),1\right]$, subject to the boundary conditions $\sigma'_0(\delta)=0$ and $w_l\left(1\right)=Q_0$. Following Appendix D of MacMinn \textit{et al.} \cite{MacMinn2016}, we start by defining two indefinite integrals,
\begin{equation}
    \mathcal{I}_1\left\{\phi\right\}=\mathcal{D}\int \mathcal{M}\left(\phi\right)\dd{\sigma'_{0}}{\phi}\mathrm{d}\phi,
\end{equation}
\begin{equation}
    \mathcal{I}_2\left\{\phi\right\}=\mathcal{D}\int \left(\frac{\phi-\phi_0}{1-\phi}\right) \mathcal{M}\left(\phi\right)\dd{\sigma'_{0}}{\phi}\mathrm{d}\phi.
\end{equation}
For the mobility law $\mathcal{M}=\phi$ and the elasticity law defined in Equation (\ref{1D elastic law}), these integrals can be evaluated exactly:
\begin{multline}
    \mathcal{I}_1\left\{\phi\right\}=\left(\frac{1+\varsigma}{1-\phi_0}\right)\frac{\phi^2}{2}+\left(1-\phi_0\right)\left[\frac{\phi}{1-\phi}+\log{\left(1-\phi\right)}\right]\\
    -\frac{\varsigma}{4\left(1-\phi_0\right)}\left[\phi^2+2\phi+2\log{\left(1-\phi\right)}-2\phi^2\log{\left(\frac{1-\phi}{1-\phi_0}\right)}\right],
\end{multline}
\begin{multline}
    \mathcal{I}_2\left\{\phi\right\}=-\left(\frac{\phi_0}{1-\phi_0}\right)\mathcal{I}_1\left\{\phi\right\}+\left(\frac{1+\varsigma}{\left(1-\phi_0\right)^2}\right)\frac{\phi^3}{3}+\left[\phi+\frac{1}{1-\phi}+2\log{\left(\phi-1\right)}\right]\\
    -\frac{\varsigma}{18\left(1-\phi_0\right)^2}\left[6\phi+3\phi^2+2\phi^3+6\phi^3\log{\left(\frac{\phi_0-1}{\phi-1}\right)}+6\log{\left(1-\phi\right)}\right].
\end{multline}
Using Equation (\ref{SI poroelasticity}) and the definitions of $\mathcal{I}_1$ and $\mathcal{I}_2$, it can be shown that
\begin{equation}\label{analyticphi1}
    \mathcal{I}_2\left\{\phi\left(1\right)\right\}-\mathcal{I}_1\left\{\phi\left(1\right)\right\}+\mathcal{I}_1\left\{\phi\left(\delta\right)\right\}-\mathcal{I}_2\left\{\phi\left(\delta\right)\right\}=Q_0\mathcal{D}.
\end{equation}
The boundary condition that the packing is stress-free at the left boundary becomes $\phi\left(\delta\right)=\phi_0$, and Equation (\ref{analyticphi1}) then becomes an implicit expression for $\phi\left(1\right)$. It can also be shown that
\begin{equation}
    \delta=\frac{1}{Q_0\mathcal{D}}\left[\mathcal{I}_2\left\{\phi\left(1\right)\right\}-\mathcal{I}_2\left\{\phi_0\right\}\right],
\end{equation}
which gives an explicit expression for $\delta$. Finally, 
\begin{equation}\label{analyticphix}
    \mathcal{I}_1\left\{f\left(x\right)\right\}-\mathcal{I}_1\left\{\phi_0\right\}=Q_0\mathcal{D}\left(\delta-x\right)
\end{equation}
gives an implicit expression for $f\left(x\right)$, which can be solved for numerically.

\section{Linear Stability Analysis}\label{LSA Details}

In \S \ref{LSA section}, we introduce the functions $h_{\alpha\beta}$ and the parameters $\nu_{\alpha\beta}$ as coefficients in our linearised evolution equations (Equations \ref{Linearised equations}). To find explicit forms for these coefficients, we first define $h_\alpha$ as the  homogeneous part of the chemical potential of fluid $\alpha$, given by
\begin{equation}
    h_g=\log{\phi_g}+1-\phi+\left(1-\phi_g\right)\left[\chi_{gl}\left(\phi-\phi_g\right)+\chi_{gs}\left(1-\phi\right)\right]+{\sigma}'_{xx},
\end{equation}
and
\begin{equation}
    h_l=\log{\left(\phi-\phi_g\right)}+1-\phi+\phi_g\left[\chi_{gl}\phi_g-\chi_{gs}\left(1-\phi\right)\right]+{\sigma}'_{xx}.
\end{equation}
To derive $h_{\alpha\beta}$, we then differentiate these expressions with respect to $\phi_g$ and $\phi$, and evaluate them at $\phi=\phi'_0$ and $\phi_g=\phi_{g0}$. This gives
\begin{subequations}
\begin{equation}
    h_{gg}=\frac{1}{\phi_{g0}}-\left(1+\phi'_0-2\phi_{g0}\right)\chi_{gl}-\left(1-\phi'_0\right)\chi_{gs},
\end{equation}
\begin{equation}
    h_{gf}=-1+\left(1-\phi_{g0}\right)\left[\chi_{gl}-\chi_{gs}\right]+\frac{1}{2\mathcal{D}}\left(2+\varsigma\right)\left(\frac{1}{1-\phi'_0}+\frac{1}{1-\phi_0}\right),
\end{equation}
\begin{equation}
    h_{lg}=-\frac{1}{\phi'_0-\phi_{g0}}+2\phi_{g0}\chi_{gl}-\left(1-\phi'_0\right)\chi_{gs},
\end{equation}
\begin{equation}
    h_{lf}=\frac{1}{\phi'_0-\phi_{g0}}-1+\phi_{g0}\chi_{gs}+\frac{1}{2\mathcal{D}}\left(2+\varsigma\right)\left(\frac{1}{1-\phi'_0}+\frac{1}{1-\phi_0}\right).
\end{equation}
\end{subequations}
To find explicit forms for $\nu_{\alpha\beta}$, we collect the coefficients of fourth order derivatives to give
\begin{subequations}
\begin{equation}
    \nu_{gg}=-\left(1-\phi_{g0}\right)\Gamma_1-\left(\phi'_0-\phi_{g0}\right)\Gamma_2-\left(2\phi_{g0}-\phi'_0-1\right)\frac{\Gamma_3}{2},
\end{equation}
\begin{equation}
    \nu_{gf}=-\left(\phi_{g0}-\phi'_0\right)\Gamma_2-\left(1-\phi_{g0}\right)\frac{\Gamma_3}{2},
\end{equation}
\begin{equation}
    \nu_{lg}=\phi_{g0}\Gamma_1-\left(\phi'_0-\phi_{g0}-1\right)\Gamma_2-\left(2\phi_{g0}-\phi'_0+1\right)\frac{\Gamma_3}{2},
\end{equation}
\begin{equation}
    \nu_{lf}=-\left(\phi_{g0}-\phi'_0+1\right)\Gamma_2+\phi_{g0}\frac{\Gamma_3}{2}.
\end{equation}
\end{subequations}
The dispersion relation derived in \S \ref{LSA section} (Equation \ref{Dispersion}) is written compactly in terms of the functions $\zeta_1\left(k\right)=a_1+a_2k^2$ and $\zeta_2\left(k\right)=b_1+b_2k^2+b_3k^4$, where 
\begin{subequations}
\begin{equation}
    a_1=M_1h_{gg}+M_2h_{lg}+M_3h_{gf}+M_4h_{lf},
\end{equation}
\begin{equation}
    a_2=-M_1\nu_{gg}-M_2\nu_{lg}-M_3\nu_{gf}-M_4\nu_{lf},
\end{equation}
\end{subequations}
and
\begin{subequations}
\begin{equation}
    b_1=\left(M_1M_4-M_2M_3\right)\left(h_{gg}h_{lf}-h_{gf}h_{lg}\right),
\end{equation}
\begin{equation}
    b_2=\left(M_1M_4-M_2M_3\right)\left(h_{lg}\nu_{gf}+h_{gf}\nu_{lg}-h_{gg}\nu_{lf}-h_{lf}\nu_{gg}\right),
\end{equation}
\begin{equation}
    b_3=\left(M_1M_4-M_2M_3\right)\left(\nu_{gg}\nu_{lf}-\nu_{lg}\nu_{gf}\right).
\end{equation}
\end{subequations}
We identify $M_i$,
\begin{subequations}
\begin{equation}
    M_1=\mathcal{M}_r\left(1-\phi_{g0}\right)\phi_{g0},
\end{equation}
\begin{equation}
    M_2=-\left(\phi'_0-\phi_{g0}\right)\phi_{g0},
\end{equation}
\begin{equation}
    M_3=\mathcal{M}_r\left(1-\phi'_{0}\right)\phi_{g0},
\end{equation}
\begin{equation}
    M_4=\left(1-\phi'_0\right)\left(\phi'_0-\phi_{g0}\right),
\end{equation}
\end{subequations}
as linearised mobility coefficients.

\bibliography{References}

\end{document}